\documentclass{iopart}
\usepackage{graphicx}
\usepackage[abbr]{harvard}
\begin{document}

\def\aj{AJ}%
\def\actaa{Acta Astron.}%
\def\araa{ARA\&A}%
\def\apj{ApJ}%
\def\apjl{ApJ}%
\def\apjs{ApJS}%
\def\ao{Appl.~Opt.}%
\def\apss{Ap\&SS}%
\def\aap{A\&A}%
\def\aapr{A\&A~Rev.}%
\def\aaps{A\&AS}%
\def\azh{AZh}%
\def\baas{BAAS}%
\def\bac{Bull. astr. Inst. Czechosl.}%
\def\caa{Chinese Astron. Astrophys.}%
\def\cjaa{Chinese J. Astron. Astrophys.}%
\def\icarus{Icarus}%
\def\jcap{J. Cosmology Astropart. Phys.}%
\def\jrasc{JRASC}%
\def\mnras{MNRAS}%
\def\memras{MmRAS}%
\def\na{New A}%
\def\nar{New A Rev.}%
\def\pasa{PASA}%
\def\pra{Phys.~Rev.~A}%
\def\prb{Phys.~Rev.~B}%
\def\prc{Phys.~Rev.~C}%
\def\prd{Phys.~Rev.~D}%
\def\pre{Phys.~Rev.~E}%
\def\prl{Phys.~Rev.~Lett.}%
\def\pasp{PASP}%
\def\pasj{PASJ}%
\def\qjras{QJRAS}%
\def\rmxaa{Rev. Mexicana Astron. Astrofis.}%
\def\skytel{S\&T}%
\def\solphys{Sol.~Phys.}%
\def\sovast{Soviet~Ast.}%
\def\ssr{Space~Sci.~Rev.}%
\def\zap{ZAp}%
\def\nat{Nature}%
\def\iaucirc{IAU~Circ.}%
\def\aplett{Astrophys.~Lett.}%
\def\apspr{Astrophys.~Space~Phys.~Res.}%
\def\bain{Bull.~Astron.~Inst.~Netherlands}%
\def\fcp{Fund.~Cosmic~Phys.}%
\def\gca{Geochim.~Cosmochim.~Acta}%
\def\grl{Geophys.~Res.~Lett.}%
\def\jcp{J.~Chem.~Phys.}%
\def\jgr{J.~Geophys.~Res.}%
\def\jqsrt{J.~Quant.~Spec.~Radiat.~Transf.}%
\def\memsai{Mem.~Soc.~Astron.~Italiana}%
\def\nphysa{Nucl.~Phys.~A}%
\def\physrep{Phys.~Rep.}%
\def\physscr{Phys.~Scr}%
\def\planss{Planet.~Space~Sci.}%
\def\procspie{Proc.~SPIE}%
\let\astap=\aap
\let\apjlett=\apjl
\let\apjsupp=\apjs
\let\applopt=\ao

\title[The nuclear cluster of the Milky Way]{The nuclear cluster of the Milky Way: Our primary testbed for the interaction of a dense star cluster with a  massive black hole.}

\author{R Sch{\"o}del$^{1}$, A Feldmeier$^{2}$, N Neumayer$^{3,2}$, L Meyer$^{4}$ and S Yelda$^{4}$}

\address{$^{1}$ Instituto de Astrof\'isica de Andaluc\'ia (CSIC), Glorieta de la Astronom\'ia s/n, 18008 Granada, Spain}\ead{rainer@iaa.es}
\address{$^{2}$ European Southern Observatory, Karl-Schwarzschild-Strasse 2, D-85748 Garching bei M\"unchen, Germany}
\address{$^{3}$ Max-Planck-Institut fuer Astronomie, K\"onigstuhl 17, D-69117 Heidelberg, Germany}
\address{$^{4}$ Department of Physics and Astronomy, University of California Los Angeles, Los Angeles, CA 90095-1547, USA}

\begin{abstract} 

This article intends to provide a concise overview, from an observational point-of-view, of the current state of our knowledge of the most relevant properties of the Milky Way's nuclear star cluster (MWNSC). The MWNSC appears to be a typical specimen of nuclear star clusters, which are found at the centers of the majority of all types of galaxies. Nuclear clusters represent the densest and most massive stellar systems in the present-day Universe and frequently coexist with central massive black holes. They are therefore of prime interest for studying stellar dynamics and the MWNSC is the only one that allows us to obtain data on milli-parsec scales. After  discussing the main observational constraints, we start with a description of  the overall structure and kinematics of the MWNSC, then focus on a comparison to extragalactic systems, summarize the properties of the young, massive stars in the immediate environment of the Milky Way's central black hole, Sagittarius\,A*, and finally focus on the dynamics of stars orbiting the black hole at distances of  a few to a few tens of milli parsecs. 

 \end{abstract}

\submitto{\CQG}

\section{Introduction}

The study of nuclear star clusters (NSCs) is a relatively young field because the necessary high angular resolution required ($\theta\approx0.1"$) has only been available since the advent of the {\it Hubble Space Telescope} (HST). NSCs have been found at the photometric and dynamical centers of about 75\% of all galaxies in the local Universe \citeaffixed{Boker:2002kx,Carollo:1998fk,Cote:2006eu,Neumayer:2011uq}{e.g.}. With effective radii of a few parsecs and masses ranging between a few times $10^{6}$ to $10^{8}$\,M$_{\odot}$, they are among the densest known stellar structures \cite{Boker:2004oq,Walcher:2005ys}. NSCs possess complex stellar populations and show clear signs of recurrent star formation, with the most recent event having occurred less than 100\,Myr ago in many of them \cite{Rossa:2006zr,Seth:2006uq,Walcher:2006ve}. An increasing number of observations show that NSCs can coexist with massive black holes (MBHs) at their centers \cite{Seth:2008rr,Graham:2009lh,Neumayer:2012fk}.

The high stellar masses and densities of NSCs as well as their potential interaction with MBHs mean that these objects are of great interest for the investigation of N-body dynamics and tests of General Relativity, as well as for studies of rare phenomena, such as tidal disruptions and stellar collisions. Some of those scientific questions  are, for example, the formation of a stellar cusp around a MBH \citeaffixed{Bahcall:1977ys,Lightman:1977ly,Murphy:1991zr}{e.g.}, resonant relaxation \citeaffixed{Rauch:1996kx,Hopman:2006fk}{e.g.,}, strong mass segregation \citeaffixed{Alexander:2009gd,Preto:2010kx}{e.g.,}, anomalous diffusion \cite{Bar-Or:2013vn}, the Schwarzschild barrier \cite{Merritt:2011ys,Hamers:2014zr,Bar-Or:2014uq}, the evolution of the orbit of a star in the immediate vicinity of  the MBH \citeaffixed{Rubilar:2001fk,Weinberg:2005fk}{e.g.},  or the study of extreme mass ratio inspiral events (EMRIs), where gravitational radiation is released during the infall of a stellar-mass object into the MBH \citeaffixed{Amaro-Seoane:2007ve}{e.g.}.

\begin{center}
\begin{figure}
\includegraphics[width=\textwidth]{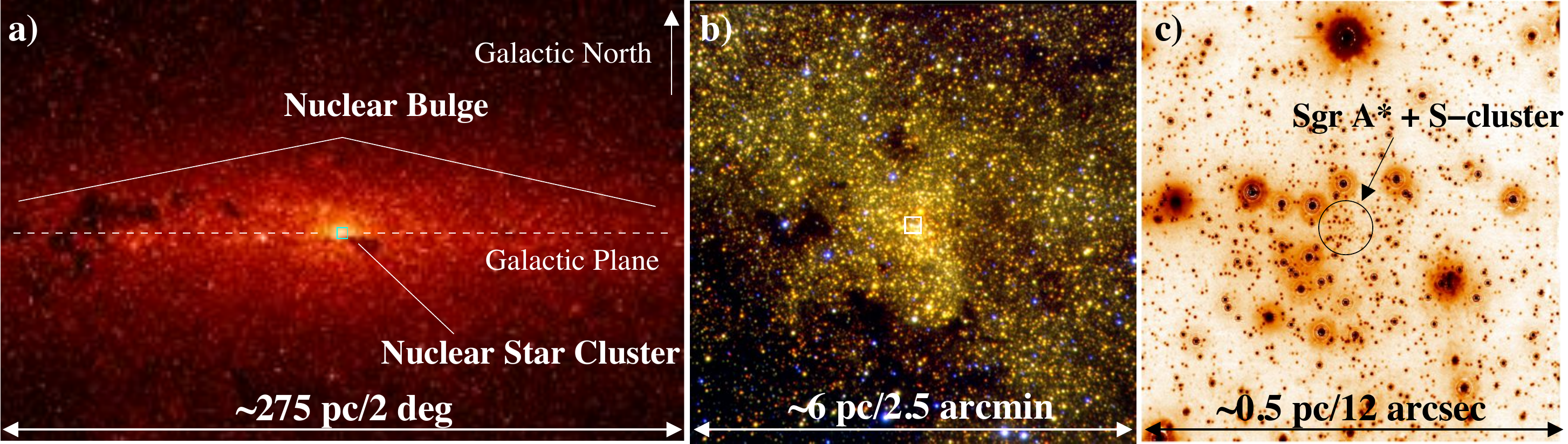}
\caption{\label{Fig:overview}  a) Extinction-corrected  $4.5\,\mu$m Spitzer/IRAC image of the GC \cite{Schodel:2014fk}. (b) ISAAC multi-color ($1.19+1.71+2.25\,\mu$m)
close-up of the field marked by the small cyan square in (a), based upon the data described in \citeasnoun{Nishiyama:2013fk}.  (c) Holographic image of the central $0.5$\,pc of the NSC (white box in (b)) from NACO/VLT \cite[resolution $\approx0.06”/2.4$\,mpc]{Schodel:2013fk}. Sagittarius\,A* is located at the center of all images (but is extremely faint at these wavelengths).}
\end{figure}
\end{center}

Since the Galactic Center (GC) is located at a distance of only about $8.0\pm0.25$\,kpc \cite{Malkin:2013fk}, high angular resolution observations with the HST or with large ground-based, adaptive-optics (AO) assisted telescopes allow us to resolve physical scales on the order of a few milli-parsecs (mpc) and thus to study the properties, kinematics and even dynamics of {\it individual} stars. At the typical near-infrared (NIR) observing wavelength of $2.2\,\mu$m (the ``K-band''), the currently best achievable angular resolution on ground-based 8-10 m telescopes with adaptive optics (AO) systems, like the ESO VLT, Gemini, or the W.M. Keck telescopes, is on the order of $0.05"$. This corresponds to about 2\,mpc, or 400\,AU, at the distance of the GC. The relative positions of stars can usually be measured to a much higher precision, typically at least a few 10 times better, depending on the brightness of the star. For intermediate-bright (with magnitudes of $K\approx14-15$) stars at the GC the astrometric precision is typically a few $0.1$ milli-arcseconds (mas) or, correspondingly, a few $0.004$\,mpc \cite{Fritz:2010fk,Yelda:2010sl}. This situation is fundamentally different from extragalactic NSCs, where we can at best study the light averaged over scales from a few $0.1$ to several pc. Finally, we possess strong observational evidence from stellar dynamics for the existence of a $4\times10^{6}$\,M$_{\odot}$ massive black hole (MBH) at the heart of the Milky Way's (MW) NSC \citeaffixed{Ghez:2003fk,Schodel:2003qp,Gillessen:2009qe,Genzel:2010fk,Meyer:2012fk}{e.g.}. 

\begin{center}
\begin{figure}
\includegraphics[width=\textwidth]{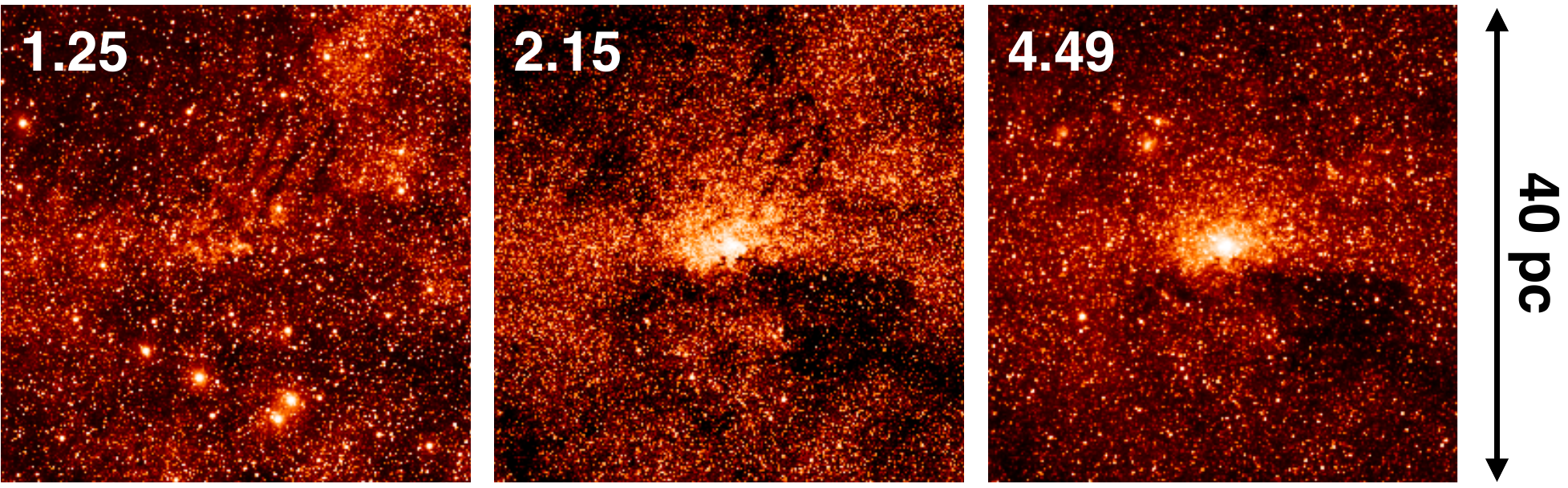}
\caption{\label{Fig:Ext}  The GC observed at wavelengths of $1.25\,\mu$m, $2.15\,\mu$m, and $4.49\,\mu$m. Sgr\,A* is located at the center of the images. The Galactic Plane runs horizontally through the middle of the images, with Galactic north up. The $1.25\,\mu$m and $2.15\,\mu$m images were taken with ESO VISTA/VIRCAM, as part of the VVV survey \cite{Minniti:2010fk}. The $4.49\,\mu$m image is  from the Spitzer/IRAC survey of the GC \cite{Stolovy:2006fk}. }
\end{figure}
\end{center}

The MWNSC is therefore a unique laboratory to test our hypotheses and theories about the properties and evolution of dense stellar systems. Here, we intend to provide an up-to-date review of our knowledge on the MWNSC and its relation to extragalactic NSCs. We will also highlight several specialized topics that have recently drawn much theoretical and observational attention, such as the dynamics of the young stars in the central parsec, the orbits of short-period stars around the MBH, and the prospects for tests of General Relativity (GR) in the MWNSC laboratory.

Figure\,\ref{Fig:overview} provides an overview of the Galactic Center (GC). The GC is outlined by the so-called nuclear bulge \citeaffixed{Launhardt:2002nx}{NB,}. The NB lies deeply embedded within the kpc-scale Galactic Bulge/Bar and is a flattened, possibly disk-like stellar structure with a radius of ~230 pc and a scale height of $\sim$45\,pc. At the heart of the NB lies the nuclear star cluster with the central black hole, Sagittarius\,A* (Sgr\,A*). Sgr\,A* is surrounded by a cluster of massive, young stars (see Section\,\ref{sec:young_stars}) and, on the smallest scales,  by a cluster of B-type dwarfs, the so-called S-stars.  The orbital motion of the S-stars around Sgr\,A* has been observed since the early 1990s and may provide key tests of GR (see Section\,\ref{sec:GR}).

\section{Observational constraints}

In spite of its outstanding importance, there exist fundamental limitations on our possibilities to study the MWNSC observationally. Knowledge of the observational constraints is important in order to understand why progress is sometimes slow and why our knowledge is still incomplete, on the one hand, and to assess the relevance and potential biases of published work, on the other hand. The two main limiting factors in studying the MWNSC are interstellar extinction and  crowding.

The line-of-sight toward the GC traverses spiral arms and the dense central molecular zone \citeaffixed{Morris:1996vn}{see, e.g.,} of the Galaxy, resulting in extreme reddening and extinction. In the NIR, the wavelength dependence of the interstellar extinction toward the GC is approximately $A_{\lambda}\propto\lambda^{-2}$ \citeaffixed{Nishiyama:2006tx,Gosling:2009kl,Schodel:2010fk}{e.g.}, where the extinction, $A_{\lambda}$, is in units of magnitudes. With $A_{1.3\,\mu m}\approx7.0$, $A_{2.2\,\mu m}\approx2.5$, and $A_{4.5\,\mu m}\approx0.5$, the corresponding attenuation factors are on the order of $0.001$, $0.1$, and $0.6$ at the respective wavelengths. At optical wavelengths the attenuation is $\sim$$10^{-12}$, rendering observations of the GC in the short NIR to UV regimes all but impossible. We illustrate this point in Fig.\,\ref{Fig:Ext}: At $1.25\,\mu$m the GC is hardly visible and most observed sources lie in the foreground, while the MWNSC is prominent at $4.5\,\mu$m. 

\begin{center}
\begin{figure}
\includegraphics[width=\textwidth]{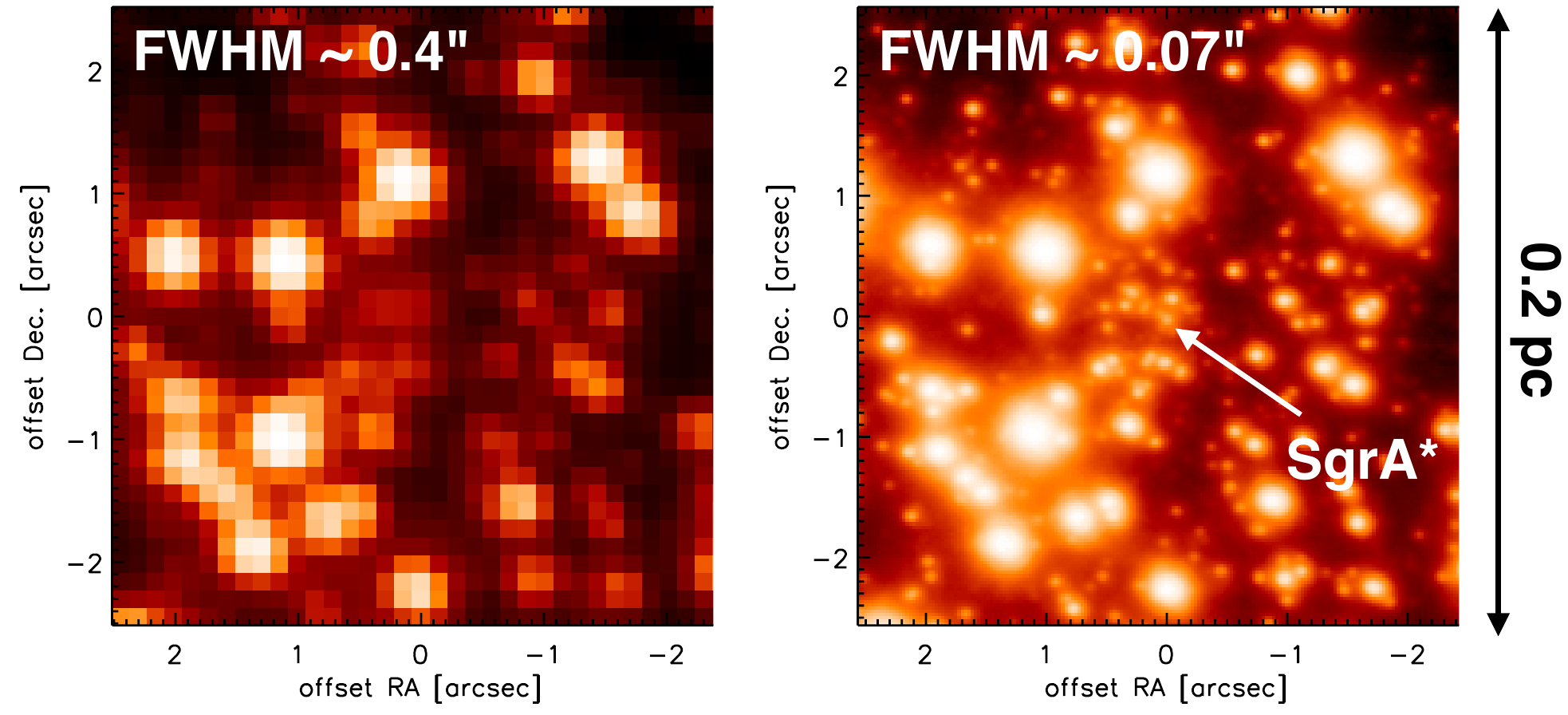}
\caption{Left: Image of the central $0.2$\,pc of the GC under excellent natural seeing \citeaffixed{Nishiyama:2013fk}{ISAAC/VLT, see}. Right: The same field observed with AO \citeaffixed{Schodel:2010fk}{NACO/VLT, see, e.g.,}. Both images are taken at $\lambda\approx2.2\,\mu$m. North is up and east is to the left.}
\label{Fig:crowding}
\end{figure}
\end{center}

The very high extinction not only poses a formidable challenge to the investigation of the large-scale properties of the MWNSC, such as its size and shape, but also to the study of its stellar population because (a) intrinsic stellar colors are small in the NIR at $\lambda>1.5\,\mu$m (not more than a few tenths of magnitudes), where most of the sensitive imaging must necessarily be done, and (b) the observed colors are dominated by the reddening due to extinction. For example, red clump (RC) giants are tracers of the old population, while B main sequence (MS) stars trace recent (a few 100 Myr) star formation.  They have both an observed magnitude of $[2.2\,\mu m]\approx15$. Their intrinsic color difference is only $[1.7\mu m]-[2.2\,\mu m]=0.3$, but the corresponding mean reddening between these wavelengths is about $2.0$. The fact  that the interstellar extinction toward the GC is also highly inhomogeneous and can vary by up to several magnitudes on spatial scales as small as a few arcseconds adds additional complication to photometric investigations of the stars at the GC. This means that detailed studies of individual stars have, so far, been largely limited to spectroscopy of the brightest sources. Spectroscopy of faint sources requires high-angular resolution and an integral field spectrometer, techniques which allow us to study only very small fields, of a few arcseconds across, but are inadequate to study the stars of the  MWNSC on large scales.

The extreme source crowding toward the GC requires an angular resolution of at least $\theta\approx0.2"$, and even higher for the immediate environment of Sagittarius\,A*. Even with AO-assisted imaging on 10m-class ground based telescopes, source confusion is one of the principal problems in studying the properties and dynamics of stars in the central arcsecond (40\,mpc) around the MBH \citeaffixed{Ghez:2008fk,Gillessen:2009qe}{see, e.g., discussion in}. Crowding is an important limitation on photometric and astrometric accuracy in the GC. Unfortunately, high-angular resolution observations require costly AO facilities and current technology can provide accurate image correction only over relatively small fields, not much larger than about one arcminute across. We illustrate the effect and importance of high angular resolution in Fig.\,\ref{Fig:crowding}, where we compare two images of the central $0.2\times0.2$\,pc$^{2}$. Both are taken with ground based telescopes, one under very good atmospheric seeing and the other one with adaptive optics (AO). Finally, we note that for most astrometric and spectroscopic work, the GC is typically observed at wavelengths around $2.2\,\mu$m. In this regime, the HST can only offer an angular resolution around $0.25"$. This is insufficient to investigate the dynamics of stars in the central $\sim$0.1\,pc near the MBH. Therefore, AO-assisted ground-based observations at 8-10m telescopes are required for this kind of work.

Together, crowding and extinction pose serious constraints on the existing observational capabilities. Because of these difficulties, we can currently only assign very rough spectral types, mostly based on luminosity, to the vast majority of stars observed at the GC. Without costly AO-assisted spectroscopy, we can, for example, usually not distinguish between a RC star and an O/B main sequence star at the GC. The faintest main sequence stars that can be detected (but not identified as such) with current facilities at a reasonable completeness in the central parsecs of the GC have at least 2 solar masses \citeaffixed{Schodel:2007tw}{see, e.g., Fig.\,16 in}. Their corresponding magnitude at $2.2\,\mu$m is about 18. A solar-mass star at the GC would have an observed magnitude in the K-band (i.e., around $2.2\,\mu$m) of approximately 21. Therefore, we will need the next generation of 30-40m-class telescopes  to study the distribution of solar mass stars in the central parsecs of the MWNSC.  Finally, we note that objects such as white dwarfs will be out of reach even for these future extremely large telescopes because they are both intrinsically faint and emit little radiation at NIR wavelengths.

\section{Overall properties of the MWNSC} 

The MWNSC was discovered by the pioneering infrared observations of \citeasnoun{Becklin:1968nx}. It was found to be an extended structure of about $5'$ diameter and elongated along the Galactic Plane (GP), with a mean projected surface brightness profile proportional to $R^{-0.8\pm0.1}$, where $R$ is the projected distance from the center, and an estimated mass of $\sim$$2.25\times10^{7}$\,M$_{\odot}$ within 5\,pc of the center. With the advent of ever larger and more sensitive infrared detectors, the  observational data improved in the  following decades considerably. The original conclusions of \citeasnoun{Becklin:1968nx} were confirmed by all follow-up observations. The perhaps most complete work that addresses the stellar structures at the GC from sub-pc scales to a few 100 pcs, was presented by \citeasnoun{Launhardt:2002nx}. They described the MWNSC as a spherically symmetric cluster of approximately $(3.0\pm1.5)\times10^{7}$\,M$_{\odot}$, embedded in a massive, disk-like structure of about $10^{9}$\,M$_{\odot}$ with a scale length of $\sim$$120$\,pc, the so-called Nuclear Stellar Disk, that is aligned with the GP (see Fig.\,\ref{Fig:overview}). 

A fundamental problem that continued to be unresolved, however,  was the question whether the MWNSC is intrinsically spherically symmetric or not and what its intrinsic size and shape is. While the MWNSC appears to be clearly flattened along the GP, this was attributed in part, or fully, due to effects of differential extinction (see previous section and Fig.\,\ref{Fig:Ext}). Spherical symmetry was generally assumed for the MWNSC to facilitate quantitative estimates and multi-particle simulations. This assumption appeared to be justified because the existing high-angular resolution data on the central parsec of the MWNSC were considered to be consistent with this assumption \citeaffixed{Schodel:2007tw,Trippe:2008it}{e.g.}. 

\begin{center}
\begin{figure}
\includegraphics[width=\textwidth]{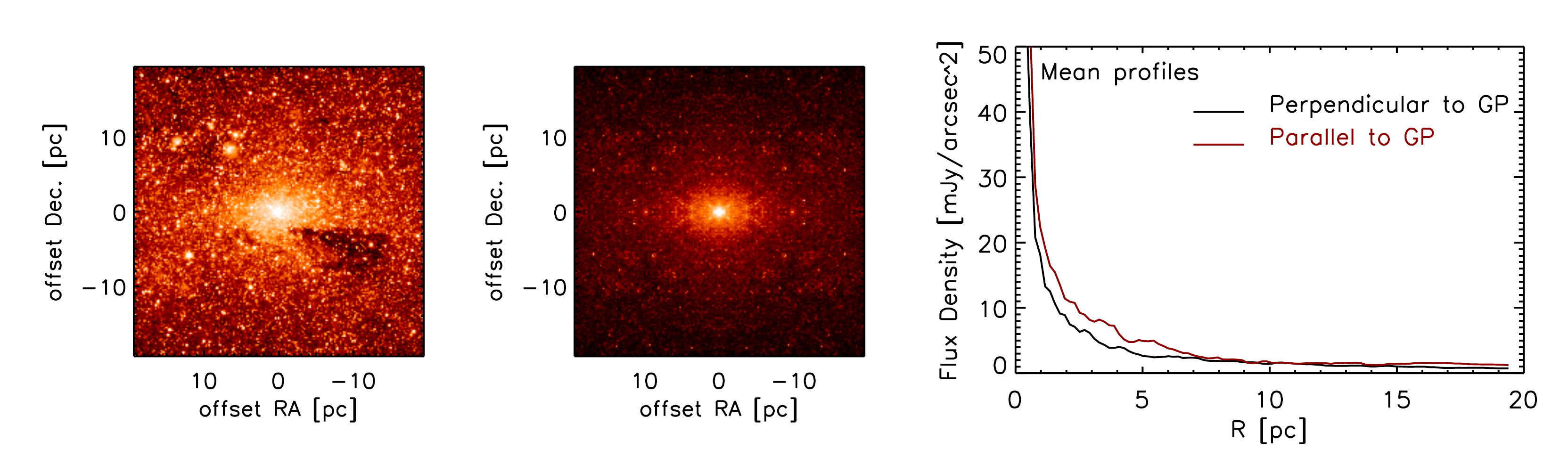}
\caption{\label{Fig:Spitzer}  Left: IRAC/Spitzer image of the MWNSC at $4.5\,\mu$m, corrected for PAH emission and extinction. Middle: Image from the left panel, after symmetrisation with respect to GP and Galactic north-south axis. Right: Light density profile of the MWNSC parallel and perpendicular to the GP. For details, see \citeasnoun{Schodel:2014fk}.}
\end{figure}
\end{center}

The high-angular resolution work by \citeasnoun{Seth:2006uq} and \citeasnoun{Seth:2008kx}, that was focused on NSCs in  nearby edge-on spiral galaxies, showed, however, that extragalactic NSCs can be flattened, which may in part be caused by their rotation. Indeed, rotation parallel to the host galaxy had also been found for the MWNSC \cite{Trippe:2008it,Schodel:2009zr}. This motivated \citeasnoun{Schodel:2014fk} to revisit the topic, using the mid-infrared (MIR) images acquired by the IRAC/Spitzer GC survey \cite{Stolovy:2006fk}. Interstellar extinction toward the GC reaches a minimum near wavelengths around $5\,\mu$m \cite{Fritz:2011fk}. By combining imaging data from IRAC Channels\,1 ($3.6\,\mu$m) and 2 ($4.5\,\mu$m), \citeasnoun{Schodel:2014fk} produced MIR images of the MWNSC that were largely corrected for extinction effects, with the exception of a few infrared dark clouds in the field-of-view. They were then able to show that, contrary to previous assumptions, the MWNSC is intrinsically elliptical and flattened along the Galactic Plane (Fig.\,\ref{Fig:Spitzer}). According to their analysis, the MWNSC 
\begin{itemize}
\item is centered on Sgr\,A* and appears point-symmetric in projection;
\item is flattened, with a  ratio between minor and major axis of $q=0.71\pm0.02$;
\item has a half-light radius of $r_h=4.2\pm0.4$\,pc;
\item has a total luminosity and mass of $L_{NSC,4.5\,\mu{m}}=4.1\pm0.4\times10^{7}\,L_{\odot}$ and $M_{MWNSC}=2.5\pm0.4\times10^{7}$\,M$_{\odot}$, respectively. 
\end{itemize}
Most of these results agree well with other, previous, work \citeaffixed{Becklin:1968nx,Launhardt:2002nx,Graham:2009lh,Schodel:2011ab,Fritz:2014vn}{e.g.} as well as with the results of kinematic modeling of the MWNSC (see section\,\ref{sec:kinematics}). But, in contrast to the previous work, \citeasnoun{Schodel:2014fk} showed, for the first time, that the MWNSC is not spherically symmetric and derived the intrinsic shape and overall properties of the MWNSC without having to rely on assumptions about the symmetry  or centering of the cluster in projection on the plane of the sky. 

\section{Does the MWNSC possess a stellar cusp?}
\label{sec:cusp}
A question of great interest for the field of stellar dynamics is the existence of a stellar cusp around Sgr\,A*. The formation of a stellar cusp around a MBH in a dense stellar cluster is a firm prediction of theoretical stellar dynamics \citeaffixed{Bahcall:1977ys,Lightman:1977ly,Murphy:1991zr,Preto:2010kx}{e.g.}. However, since the physical scale of cusps is on the order of the radius of influence of the central black hole, i.e.\ the radius that contains about twice as much mass in stars as in the black hole \citeaffixed{Merritt:2010ve}{e.g.,}, their observable  {\it angular} scales are usually below the resolution power of existing telescopes in the case of extragalactic NSCs. In addition, we can only observe the integrated light in extragalactic NSCs, but the latter can be easily  biased by the presence of a small number of bright stars. Therefore, the MWNSC, where we can, in fact, count {\it individual} stars within the $\sim2$\,pc radius of influence of Sgr\,A*, is a unique test case.

Theoretical work predicts that the stars in a relaxed cluster should form a power-law density distribution of the form $\rho\propto r^{-\gamma}$ around the MBH, where $\rho$ is the stellar density and $r$ is the distance form the MBH.  In the so-called weak mass segregation regime, considered relevant, e.g., for globular clusters,  the heavy stars have an $r^{-7/4}$ cusp and the light ones a shallower $r^{-3/2}$ cusp \citeaffixed{Bahcall:1977ys,Alexander:2005fk}{e.g.,}. In the strong segregation regime, which is considered more adequate for the MWNSC, the rare massive stellar objects form a $\sim$$r^{-11/4}$ cusp and the light objects a $\sim$$r^{-3/2}$  cusp \cite{Alexander:2009gd,Preto:2010kx}.  Collisions can lead to flatter cusps in high-density nuclei \citeaffixed{Murphy:1991zr}{e.g.}, but the necessary extreme densities are probably not achieved in the GC. If there exists a cusp in the GC, then collisions will only become relevant in the innermost $\sim$$0.05$\,pc \cite{Alexander:1999fk}. 

First star counts on sensitive, high-resolution AO-assisted images from 8m-telescopes indicated a density law consistent with such values, with $\gamma_{observed}=-1.4\pm0.1$ \cite{Genzel:2003it}. However, once subsequent analyses with refined methodology excluded the massive, young stars, that cannot be dynamically relaxed and thus cannot form part of the equilibrium cusp, it became evident that the stellar density within a few $\sim0.1$\,pc of Sgr\,A* is too flat to be  consistent with a classical cusp. It is important to note, however, that this statement applies only for the stars that are bright enough ($K<16$) to be classified by spectroscopic or photometric means \cite{Buchholz:2009fk,Do:2009tg,Bartko:2010fk}.

The efforts to reconcile theory with observations can be sorted grossly into the following scenarios: (1) The cusp does exist but is invisible because (a) the giant stars that are the visible tracers of the density of the relaxed cluster population have been (partly) deprived of their atmospheres and thus been rendered too faint to be observed, or (b) the cusp is composed of stellar mass black holes. (2) The cusp around Sgr\,A* has not yet been formed or has been destroyed. Scenario (1a) requires collisions to remove the large, thin envelopes of giants. Mostly, collisions between stars or between stars and stellar remnants have been considered \citeaffixed{Dale:2009ca}{e.g.}, but recently it has been suggested that collisions between giants and dense clumps of gas in a star forming disk around Sgr\,A* could have stripped the giants and thus rendered them too dim to be observable with current technology \cite{Amaro-Seoane:2014fk}. Scenario (1b) has been shown to be a plausible result of the combination of star formation and dynamical evolution of the NSC. Mass segregation may have led to the expulsion of lighter stars and the accumulation of stellar mass black holes in the central few $0.1$\,pc \cite{Lockmann:2010fk}. 
Scenario (2) would imply that the relaxation timescales in the MWNSC are longer than previously assumed  \cite{Merritt:2010ve} or that the cusp was destroyed by the infall of an intermediate-mass black hole and had not yet had the time to regrow \citeaffixed{Merritt:2006dq,Baumgardt:2006fk}{e.g.}.

The jury is still out on the question whether the MWNSC possesses a central stellar cusp as it is predicted by theoretical dynamics for a cluster relaxed by two-body interactions. Observational data on the spatial distribution of fainter, lower mass stars are needed. Correcting the star counts for systematic effects like crowding, extinction, and  classification is complex and therefore subject to potentially large uncertainties. The completeness of the identification of stellar type, for example,  depends both on the brightness of the targets and on their location within the field \citeaffixed{Bartko:2010fk,Do:2009tg}{see, e.g.,}. All we can say currently is that the distribution of the  brightest late-type stars, which make up only a few percent of the expected total stellar content of the GC, does not agree with the classical stellar cusp that is predicted to form around a MBH via two-body relaxation. The observable {\it young, massive} stars, on the other hand,  do show a cusp-like increase of their density toward Sgr\,A*, but they are too young to represent a kinematically relaxed population.

Probably, obtaining conclusive data will require the sensitivity and angular resolution of a 30-40m-class telescope. Such an instrument would allow us to detect much fainter stars in the GC and thus to obtain a more complete picture of the structure of the NSC and possible effects of mass segregation. The presence of a low-luminosity cusp or even dark cusp can be detected by dynamical effects on the observed stars, such as the requirement of an extended mass component to accurately reproduce the gravitational potential near Sgr\,A* or perturbations of stellar orbits through close encounters \citeaffixed{Weinberg:2005fk,Perets:2009qf}{e.g.,}.

\section{Kinematics of the MWNSC}  
\label{sec:kinematics}

In the stellar kinematics we can find clues about the formation of the MWNSC and its accretion history. There are two prevailing formation scenarios for NSCs. They could have formed  by infalling star clusters \citeaffixed{tremaine75,Agarwal:2011bh,Antonini:2012fk,Gnedin:2014fk}{e.g.,}, or by the accretion of gas clouds that formed the stars {\it in situ} \citeaffixed{milos04,pflamm09}{e.g.}. A combination of both scenarios is also possible and even highly probable \citeaffixed{Neumayer:2011uq,Hartmann:2011uq}{e.g.}. The two-body relaxation time of the MWNSC  is of the order of Gyr at all radii, and close to the Hubble time beyond 1\,pc \cite{Merritt:2010ve}. So  we can expect to see  an imprint of accretion events of infalling star clusters or gas clouds in the stellar kinematics, e.g.\ in the form of rotation or anisotropy. Such data could help to understand the formation and growth of galactic nuclei and of central MBHs in other galaxies as well. Knowledge of the kinematics is also the key to deriving the mass distribution of the MWNSC. 

Observationally, we can distinguish two methods to determine stellar kinematics: (1) observe stellar spectra and measure the radial velocity along the line-of-sight using spectral lines, (2) imaging stars at different epochs and measure the two-dimensional proper motion of the stars in the plane of the sky. The latter observing technique is only possible in the Galactic centre due to its close proximity. To bring those two different measurements  to  the same units, we need to infer a distance of the stars, in order to transform the observed mas/yr of the proper motion to the physical unit km/s. At a  distance of 8.0\,kpc, 1" corresponds to 0.039\,pc/8,000\,AU, and 1\,mas/yr corresponds to 38\,km/s.

Due to extinction (see Section 2),  observations are limited  either to the NIR, or to radio bands. Radial velocity measurements are often obtained in the NIR K-band and make use of the first or several of the   stellar CO absorption lines at  2.29$-$2.39\,$\mu m$\,  \citeaffixed{McGinn:1989kx,Trippe:2008it}{e.g.}, or the Na~I doublet at 2.206 and 2.209 $\mu m$ \citeaffixed{Do:2013fk}{e.g.}. Those lines are prominent in the late-type population of stars. For early-type stars, Br~$\gamma$ absorption at 2.1661\,$\mu m$\,\citeaffixed{Ghez:2008fk,Gillessen:2009nx}{e.g.} or He I lines at 2.058$-$2.113\,$\,\mu m$  \citeaffixed{Haller:1996rw,Tanner:2006vn,Zhu:2008fk}{e.g.} are used. Proper motions are measured  with    H and K band  filters  in the NIR \citeaffixed{Schodel:2009zr}{e.g.}. 
Radio observations use maser transitions of  H$_2$O \citeaffixed{Sjouwerman02}{22 GHz, e.g.}, SiO \citeaffixed{Deguchi:2004fk,Reid:2007vn}{43 GHz, e.g.}, or OH  \citeaffixed{winnberg,Lindqvist:1992ff}{1612 GHz, e.g.} to determine both, radial velocities and proper motions. 

Radial velocities  measured from masers are very accurate,  with uncertainties of only a few km/s. Maser transitions are generated in Asymptotic Giant Branch (AGB) stars, which are in the age range of  $10^7 - 10^9$\,yr \cite{Deguchi:2004fk} and spread over the MWNSC.
On the other hand,  CO bandheads, which are observed in the NIR,  have the advantage that they are not only prominent in giant and supergiant stars, but also  in dwarf stars, though weaker. The spectral types with CO bandheads range from  $\sim$G to M \cite{Wallace:1997fk}. 

Especially the NIR kinematic data sets have been focused, so far, to a region within about 1\,pc projected distance from Sgr~A*. By monitoring the centre of the MWNSC over many years,   an immense amount of stellar velocities has been obtained \cite{Trippe:2008it,Schodel:2009zr}. These data reveal that the MWNSC  is rotating in the same sense as the Galactic disk. 
Rotation implies that a nuclear star cluster accreted material from the Galactic disk \cite{Seth:2008kx}.   
 Assuming spherical symmetry, recent studies \citeaffixed{Trippe:2008it,Schodel:2009zr,Do2013td,Fritz:2014vn}{e.g.} found that the central cluster kinematics is consistent with being  isotropic (i.e. $\beta$=0). This finding  can be reconciled with the rotation of the cluster by recalling that the rotational velocity increases with distance from Sgr\,A*, while the velocity dispersion decreases. The velocity dispersion throughout most of the central parsec is significantly larger than the rotational velocity. This masks rotation in the central parsec \citeaffixed{Schodel:2009zr}{see also dicussion in}. 

 The velocity dispersion parallel to the Galactic plane is increased with respect to the velocity dispersion perpendicular to the Galactic plane \cite{Trippe:2008it,Schodel:2009zr}.  This could be caused by flattening \cite{chatzopoulos}.  While \citeasnoun{Genzel:2003it} and \citeasnoun{Trippe:2008it} suggested  that the late-type population of stars is dynamically relaxed,  \citeasnoun{Do2013td} found that the stars are unrelaxed within r\textless0.5\,pc of Sgr~A*.
Possible reasons for an unrelaxed stellar system at r$\sim$0.5\,pc   are the infall of a MBH \cite{Merritt:2010ve} or of a globular cluster \cite{Antonini:2012fk}.

Beyond the central parsec, kinematic data often come from maser stars. \citeasnoun{Lindqvist:1992ff} obtained 134 radial velocities of OH maser stars out to 100\,pc projected distance from Sgr~A*. \citeasnoun{Deguchi:2004fk} collected radial velocities of 180  maser stars in a region of $\sim56$\,pc$\times56$\,pc.  These data are often used to compute the enclosed mass of the Milky Way out to several tens of parsecs projected distance. But using less than 200 stars to trace the kinematics over a region of \textgreater 50\,pc radius implies a large uncertainty. Also,  the assumption of spherical symmetry breaks down at such large scales \cite{Launhardt:2002nx,Schodel:2014fk}.

An alternative approach to obtain radial velocity measurements beyond 1\,pc and over a large area is to observe not individual stars, but the integrated light of many stars. This was done by \citeasnoun{McGinn:1989kx}, \citeasnoun{Sellgren:1990cl}, and \citeasnoun{Feldmeier:2014kx}. \citeasnoun{McGinn:1989kx} obtained data 
out to a projected distance of 3.6\,pc, but contamination by  bright individual stars caused large scatter in the data. For that reason \citeasnoun{Sellgren:1990cl} used smaller apertures  and focused on the central parsec. 
\citeasnoun{Feldmeier:2014kx} scanned the central region of 60\,pc$^2$. 
They used star catalogs to remove foreground stars and bright individual sources from the data, thereby minimising contamination effects.

Both, \citeasnoun{McGinn:1989kx} and \citeasnoun{Feldmeier:2014kx} detected systematic rotation along the Galactic plane also for large radii.  The rotation velocity is increasing from the centre and  flattening out beyond $\sim$1.5\,pc to $\sim$40\,km/s \cite{Feldmeier:2014kx,Fritz:2014vn}.    \citeasnoun{Trippe:2008it} derived a higher rotation velocity  of $\sim$90\,km/s, but they used only a subset  of  the \citeasnoun{McGinn:1989kx} data, which  was contaminated by individual bright stars.  
\citeasnoun{Feldmeier:2014kx} found indications for an offset of the rotation axis from the photometric minor axis by $\sim$9$^\circ$ beyond $\sim$1.5\,pc projected distance from Sgr~A*. Such a position angle offset could be the signature of an infalling star cluster that is merging with the MWNSC. The velocity dispersion increases towards the centre of the MWNSC \citeaffixed{McGinn:1989kx,Schodel:2009zr}{e.g.}. Random motion  dominates over the systematic rotation component at all measured radii, especially towards the centre, where the MBH mostly influences the stellar kinematics \cite{Feldmeier:2014kx}. The specific angular momentum  parameter  $\lambda_R$ also quantifies the amount of ordered versus random motion, and its value of $\sim$0.36 indicates that  the MWNSC has similar rotational support as a fast rotating elliptical galaxy has \cite{Feldmeier:2014kx}. 
The MWNSC lies below the $\lambda_R - \epsilon$ relation for isotropic systems \cite{lambda}, where $\epsilon$ is the photometric ellipticity. This suggests that there is anisotropy in the kinematics of the MWNSC at distances $>$4\,pc from Sgr~A* \cite{Feldmeier:2014kx}.  
\begin{figure}
\begin{center}
	\includegraphics[scale=0.55]{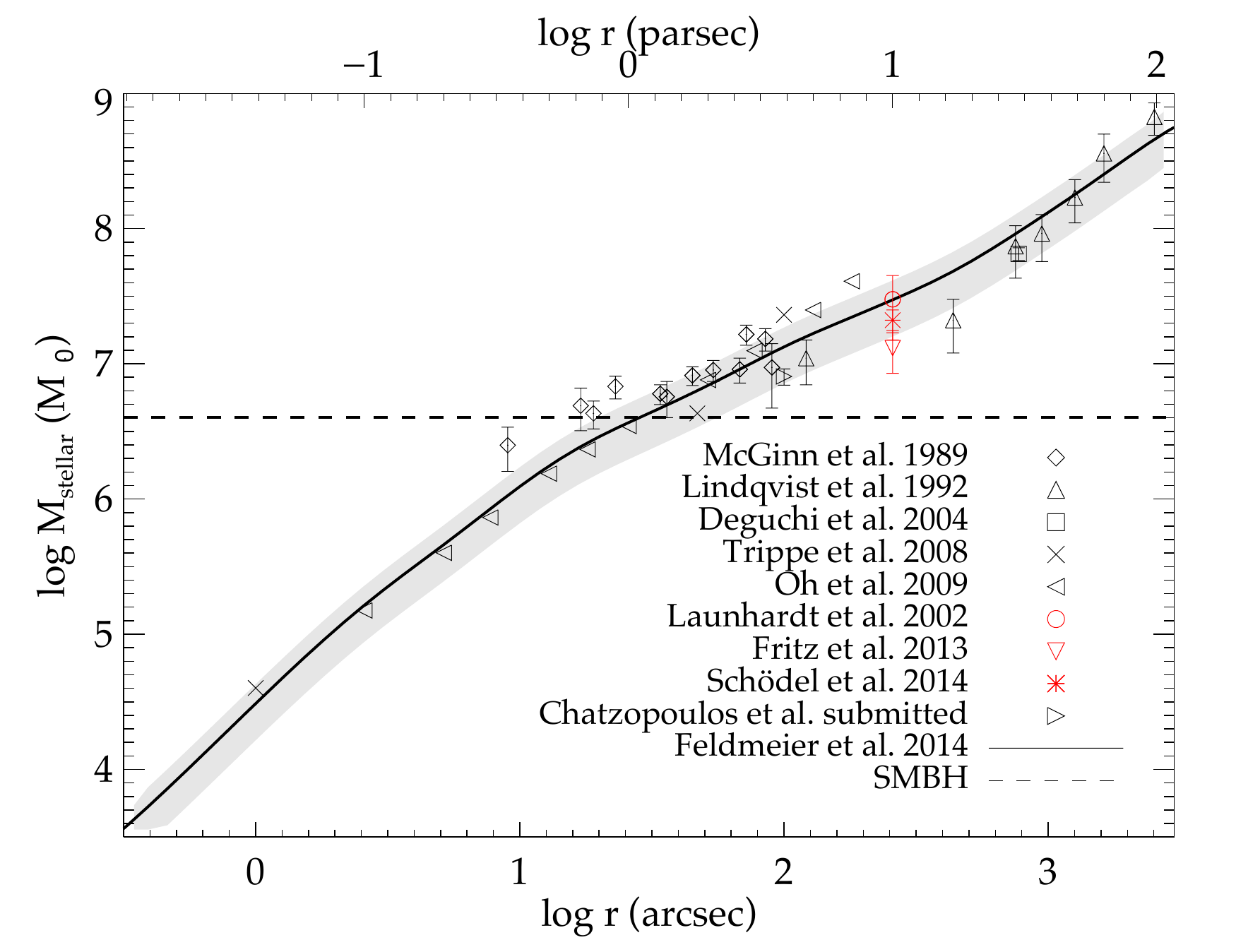} 
	\caption{Enclosed stellar mass of the MWNSC from 0.01pc to 100 pc obtained by various studies \cite{McGinn:1989kx,Lindqvist:1992ff,Deguchi:2004fk,Trippe:2008it,Oh:2009kx,Feldmeier:2014kx,chatzopoulos}. Results from studies that provide the {\it total} mass of the MWNSC \cite{Launhardt:2002nx,Fritz:2014vn,Schodel:2014fk} are all plotted in red at the same (somewhat arbitrary) distance of 10\,pc, more than twice the half-light radius, to facilitate their comparison. At this distance some of the enclosed mass can already be assigned to the nuclear stellar disk. Therefore the red points lie below the gray uncertainty band.}
	\label{fig:mass}
	\end{center}
	\end{figure}

Various of the aforementioned studies also derived a mass for the MWNSC, usually assuming spherical symmetry. Recently \citeasnoun{Schodel:2014fk} showed that the MWNSC does not appear spherical in projection, but is consistent with point-symmetry. Therefore axisymmetric models  \citeaffixed{Feldmeier:2014kx,chatzopoulos}{e.g.} are more suitable to describe the MWNSC.  Figure \ref{fig:mass} summarises the mass measurements from various studies. The enclosed stellar mass is plotted versus the distance $r$  from the center in circular apertures. In the case of \citeasnoun{Feldmeier:2014kx}, $r$ is the mean radius of elliptical apertures. These mass measurements did not strictly disentangle the MWNSC from the nuclear stellar disk, only \citeasnoun{Fritz:2014vn} and \citeasnoun{Schodel:2014fk} analysed both entities separately. 
Beside the different kinematic data, the assumed density profiles influence the results. The assumption of a constant mass-to-light ratio is often made, but as the stars in the central 0.5\,pc of the MWNSC are younger than at large radii, this may not be strictly true. But the observational difficulties (see Section 2) have, so far, impeded obtaining sufficiently complete and accurate data for such an analysis.
 Nevertheless, the different results agree quite  well. 


\section{Comparison with extragalactic nuclear star clusters} 
In this section we compare the properties of the MWNSC with those of nuclear star clusters observed at the centers of other galaxies.
The Milky Way is by far not the only galaxy to host a NSC. In fact, these are very common structural components at the centers of galaxies. 
They are found in $\sim$$77\%$ of late type galaxies \cite{Boker:2002kx,Georgiev:2014ve}, $55\%$ of spirals \cite{Carollo:1998fk}, and at least $66\%$ of (dwarf) 
ellipticals and S0s \cite{Cote:2006eu}. All of these nucleation fractions are presumably lower limits, as in some cases it is difficult to pick up the nuclear cluster due to dust \cite{Carollo:1998fk}, or  due to the high surface brightness 
of the underlying galaxy \cite{Cote:2006eu}. 

\begin{figure}[htbp] 
\begin{center} \includegraphics[width=12cm]{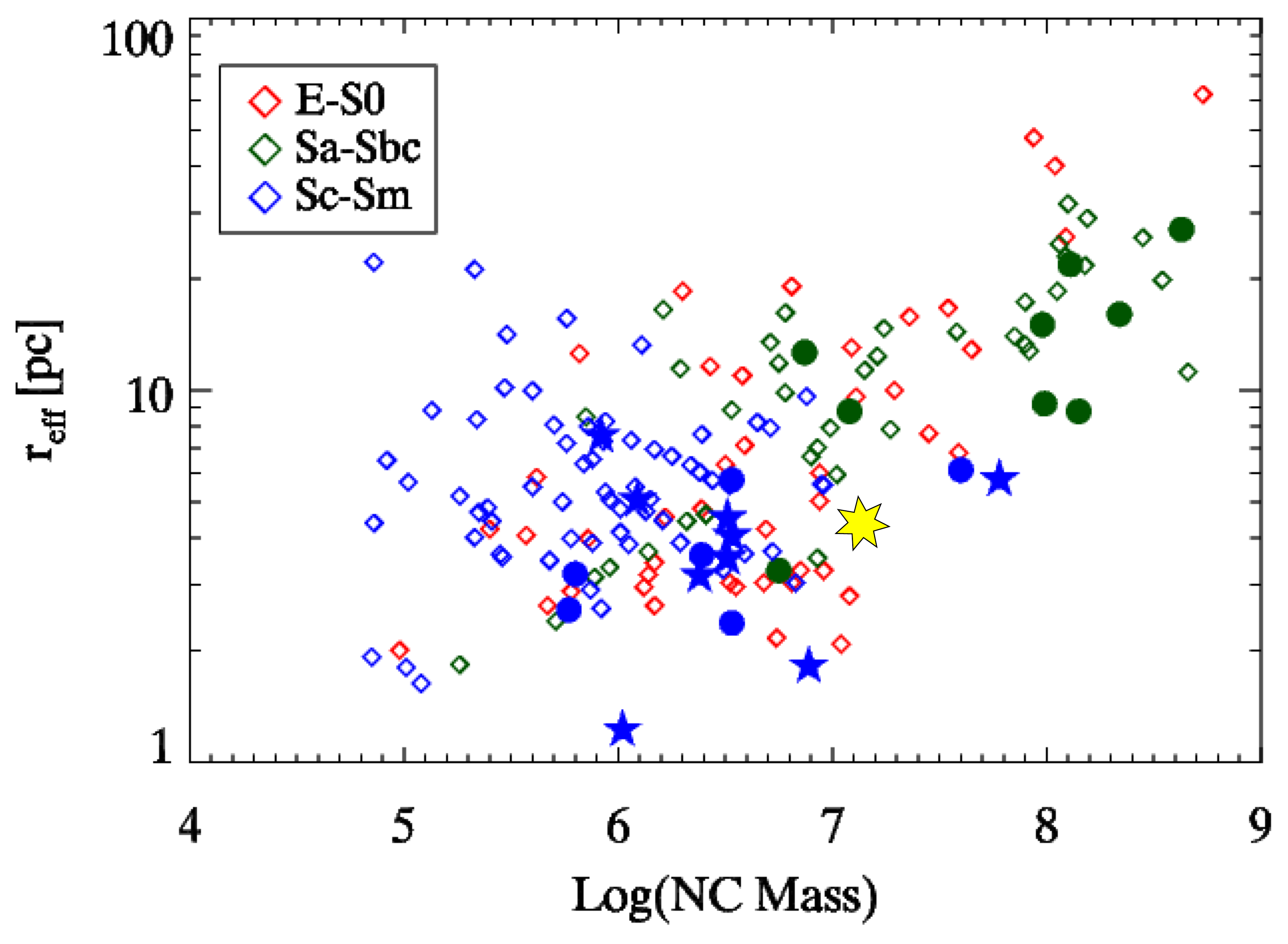}
\caption{ Nuclear cluster mass {\it vs} effective radius for nuclear star clusters (reproduced after \citeasnoun{Feldmeier:2014kx}), based on \citeasnoun{Seth:2008rr} with data from \citeasnoun{Walcher:2005ys}; \citeasnoun{Rossa:2006zr} shown as filled symbols and \citeasnoun{Boker:2002kx}; \citeasnoun{Carollo:1997ys}, \citeasnoun{Carollo:1998fk},\citeasnoun{Carollo:2002uq}, \citeasnoun{Cote:2006eu} as open symbols.  The yellow star is the result for the MWNSC using a half-light radius of 4.2pc \cite{Schodel:2014fk}.}  \label{mass-size} 
\end{center} 
\end{figure}

Typical half light radii of nuclear clusters range between $3-5$pc in late type spiral galaxies \cite{Boker:2002kx,Boker:2004oq,Georgiev:2014ve}, but can be up to several tens of parsec in earlier type galaxies \cite{Carollo:1998fk,Cote:2006eu}. These larger sizes measured in earlier galaxy types may be partly due to confusion with a surrounding disk, such as the nuclear stellar disc at the center of the Milky Way \cite{Launhardt:2002nx},  or contamination from the bulge. The range for spiral galaxies perfectly brackets the value of 4.2pc that \citeasnoun{Schodel:2014fk} find for the MWNSC, which is in excellent agreement with the value of 4.4pc, recently published by \citeasnoun{Fritz:2014vn}. These sizes are comparable to Milky Way globular clusters. However, NSCs are on average 4mag brighter than globular clusters, with total I-band magnitudes in the range of $-8$ to $-12$ mag \cite{Boker:2004oq}.  These high luminosities (when compared to globular clusters) are at least partially due to the higher masses of NSCs. Dynamical mass measurements derive values in the range of $1 \times 10^5 - 5\times10^8$M$_{\odot}$ \cite{Carollo:2002uq,Walcher:2005ys,Ferrarese:2006ly,Barth:2009kx,Seth:2010fk,Lyubenova:2013fk}. The mass of the MWNSC falls right at the centre of this range, making the MWNSC a typical NSC in many respects. Figure~\ref{mass-size} shows a plot of mass {\it vs} effective radius for a compilation of NSCs in different types of host galaxies.

In general, NSCs lie at the highest mass end of the star cluster mass function, and are structurally very different from bulges \cite{Walcher:2005ys,Misgeld:2011kx}.  In fact, NSCs are the densest stellar systems in the universe, with surface mass densities of typically a few $10^5$M$_{\odot}$pc$^{-2}$ up to $\sim 8 \times 10^6$ M$_{\odot}$pc$^{-2}$ \cite{Walcher:2005ys,Seth:2010fk,Misgeld:2011kx,Norris:2014uq}. For the MWNSC we get an average surface mass density of $\sim 2 \times 10^5$M$_{\odot}$pc$^{-2}$ within the effective radius of 4.2pc, and $\sim 2.5 \times 10^6$M$_{\odot}$pc$^{-2}$ within the central $\sim$$0.5$\,pc \cite{Genzel:2010fk,Feldmeier:2014kx}.

The high luminosity of nuclear clusters is not solely due to their large masses. It also results in part from the presence of young stellar populations, such as they are also observed at the GC (see section \ref{sec:young_stars}). Photometry and spectra of a number of galaxies suggest that nuclear clusters in both spiral and dwarf elliptical galaxies have populations much younger than globular clusters \citeaffixed{Ho:1995dq,Lotz:2004cr,Seth:2010fk}{e.g.}. In late-type galaxies, most clusters appear to contain stars with ages $<$100 Myr \cite{Walcher:2006ve,Seth:2006uq}. Furthermore, several spectral studies have shown that nuclear clusters are made up of composite stellar populations, with most having substantial old ($>1$ Gyr) stellar components \cite{Long:2002ve,Sarzi:2005ly,Walcher:2006ve,Rossa:2006zr,Seth:2006uq,Seth:2010fk,Pfuhl:2011uq,Lyubenova:2013zr}.

The complex star formation history of NSCs likely results from their special location in their host galaxies. For late-type spiral galaxies it has been shown both photometrically \cite{Boker:2002kx} and kinematically \cite{Neumayer:2011uq} that NSCs truly occupy their centers, meaning they sit at the bottom of the potential well. This means that gas and also young star clusters that are formed in the disk can spiral towards the center due to dynamical friction \cite{Bekki:2010qf,Agarwal:2011bh,Neumayer:2011uq,Antonini:2013ys}. Several studies have shown that NSCs have multiple stellar populations both in late type \cite{Walcher:2006ve,Rossa:2006zr,Seth:2006uq} and also early type galaxies \cite{Seth:2010fk,Lyubenova:2013fk}.  NSCs seem to be typically more metal-rich and younger than the surrounding galaxy \cite{Koleva:2011nx,Lyubenova:2013zr}. This appears also to be true for the MWNSC \citeaffixed{Ramirez:2000ys,Cunha:2007oq}{e.g.,}. In general the abundance ratios of [$\alpha$/Fe] show that NSCs are more metal enriched than globular clusters \cite{Evstigneeva:2007kl}. This finding suggests that NSCs cannot solely be the merger product of globular clusters, but need some gas for recurrent star formation. This finding is also supported by recent kinematical studies \cite{Hartmann:2011uq,De-Lorenzi:2013tg}, where cluster infall alone cannot explain the dynamical state of the NSC.

In a study of edge-on disk galaxies, \citeasnoun{Seth:2006uq} find that all of the well-resolved clusters appear flattened. Their median axis ratio (q=b/a) is 0.81, with q$\sim 0.4$ for NGC 4206 and NGC 4244, the two systems with the most prominent disks. The flattening of the MWNSC was derived by \citeasnoun{Schodel:2014fk} to be $0.71$, i.e. in the range of what has been found in other nearby edge-on galaxies.  Moreover, studies of the kinematics of the NSC in the nearby edge-on galaxy NGC\,4244 with integral-field spectroscopy show that the cluster as a whole rotates \cite{Seth:2008kx,Seth:2010fk}.  This observation is very much in-line with the kinematic study of \citeasnoun{Feldmeier:2014kx}, that shows similar kinematic structures of the MWNSC.  The misalignment of the kinematic to the photometric axis that \citeasnoun{Feldmeier:2014kx} report, is not seen in NGC\,4244. However, the position angles of the three multi-component nuclear clusters (IC\,5052, NGC\,4206, and NGC\,4244) that \citeasnoun{Seth:2006uq} studied, are all aligned within 10$\deg$ of the galaxy disk position angle.

NSCs do co-exist with black holes \cite{Seth:2008rr}. The best studied example is indeed the MWNSC. However,  there are also NSCs with very tight upper limits on the mass 
of a central black hole (see \citeasnoun{Neumayer:2012fk} for an overview). 
Combined  with the superb spatial resolution of adaptive optics, the 2D velocity maps of NSCs in nearby galaxies resolve stellar and gas kinematics down to a few parsecs on physical scales. 
In addition, due to the extremely high central stellar density in NSCs, it becomes possible to pick up kinematic signatures for black holes inside NSCs in nearby galaxies \cite{Seth:2010fk,Lyubenova:2013zr}. 

To conclude, the MWNSC appears to be similar to extragalactic NSCs in all major aspects. It may thus serve as an adequate, and unique,  template for a detailed study of the properties of NSCs.  As concerns the formation of NSCs, the work to understand the formation history of the MWNSC has only just begun. However, we can already draw some conlucions: The flattening of the MWNSC along the Galactic plane and its rotation parallel to overall Galactic rotation suggest that it formed from material -- be it gas or star clusters -- that fell in preferentially from the direction of the Milky Way's disk. 

The fact that the majority of the MWNSC's stars formed already several Gyr ago \cite{Pfuhl:2011uq} is consistent with, but does not provide any conclusive evidence for, the globular cluster infall scenario. So far, no old, low-metallicity stellar population, as it is typical for globular clusters, has been found in the MWNSC. Existing studies point to an approximately solar meallicity in the GC environment \cite{Carr:2000zr,Ramirez:2000ys,Najarro:2004kx,Davies:2009ly,Najarro:2009vn,Cunha:2007oq,Ryde:2014uq}. But,  given the observational difficulties, in particular the limitation to the near-infrared and the challenge of obtaining high resolution spectra of moderately bright giants in a very crowded field, it may well be that an old, low-metallicity population has eluded detection so far. Also, the existing metallicity studies of stars in the GC were limited to a handful of bright supergiants or very massive young stars, with only very few stars probed in the MWNSC proper. Hence, while we lack any evidence for the globular cluster infall scenario, we cannot rule it out, either. 

As concerns  growth through {\it in-situ} star formation, the young massive stars in the central parsec of the MWNSC provide  clear evidence for this mechanism. The potential presence of a coherent kinematical feature outside of the central parsec may also be interpreted as a hint that accretion of star clusters could contribute to the growth of the MWNSC \cite{Feldmeier:2014kx}. Significant observational efforts will still be needed to arrive at a clearer picture of the formation history of the MWNSC.

\section{The young, massive stars near the central black hole} 
\label{sec:young_stars}
While the bulk of the MWNSC is made up of old stars ($>$5 Gyr), there also exists an enigmatic population of nearly 200 hot, early-type stars, including Wolf-Rayet (WR) stars and O and B type main sequence stars, giants, and supergiants \cite{Allen:1990hc,Krabbe:1991ij,Krabbe:1995fk,Blum:1995qa,Tamblyn:1996mi,Najarro:1997qe,Ghez:2003fk,Paumard:2006xd,Bartko:2010fk,Pfuhl:2011uq,Do:2013fk}.  With age estimates of 3-8 Myr \cite{Paumard:2006xd,Lu:2013fk}, their presence in the central parsec ($R <$ 0.5 pc) raises the question of how stars can form in such a hostile environment, where tidal forces from the MBH will destroy typical molecular clouds before they can collapse to form stars \citeaffixed{Morris:1993ve,Genzel:2003it}{see, e.g.,discussions in}.  As discussed below, stars actually can form {\it in-situ} in the GC. Tidal forces are a problem only for gas that accumulates in one place by virtue of its own self-gravity (as is the case for giant molecular clouds that form stars in the field). The situation is different in case of a gas disk held in place by the gravity of the MBH until it accumulates enough mass to become (very briefly) self-gravitating, which is the point when it fragments and forms stars \citeaffixed{Milosavljevic:2004bh}{e.g.,}.

Clues to the origin of the young stars can be gained through detailed study of their spatial distribution and orbital dynamics. These properties should contain imprints of the stars' origin since their age is much less than the two-body relaxation timescale in the GC, which is $\mathcal{O}$(1 Gyr) \cite{Merritt:2013uq}.  Observations to date have revealed that the young stars in the central 1 pc fall into at least three distinct dynamical categories: 1) an isotropically-distributed cluster at $R <$ 0.8" (0.03 pc) consisting of primarily B-type main sequence stars with high eccentricities ($\bar{e}$ = 0.8), often referred to as the 'S-star cluster', 2) a moderately eccentric ($e \sim$ 0.3) clockwise (CW) rotating stellar disk with an inner edge at $\sim$0.8", and 3) an off-disk population also outside the central arcsecond that appears to be more isotropically distributed \cite{Genzel:2000hc,Genzel:2003it,Levin:2003kx,Paumard:2006xd,Ghez:2008fk,Lu:2009bl,Gillessen:2009qe,Bartko:2009fq,Bartko:2010fk,Yelda:2014fk}.  The stars making up the latter two groups have been shown to be coeval \cite{Paumard:2006xd}. Whether the stars in the central arcsecond are less massive members of the outer population is still an open question, although recent work has shown that the CW disk likely includes some B-type main sequence stars \cite{Yelda:2014fk,Madigan:2014pi}. If the S-stars were formed in the same starburst as the more massive stars a few Myr ago, they must have been dynamically injected into the central arcsecond, as this region is inhospitable to star formation \cite{Morris:1993ve}.  Here we review the observed properties of the stellar disk and off-disk population and discuss the implications for star formation theories.  The central arcsecond cluster will be discussed in the next section.

\subsection{Spatial Distribution and Stellar Dynamics}
\label{sec:dynamics}
The surface density profile, $\Sigma \propto R^{-\Gamma}$, of the {\it entire} known population of young stars in the central parsec is relatively steep ($\Gamma \sim$ 1) compared to that of the late-type giants in the same region ($\Gamma \sim$ 0; see \S\ref{sec:cusp}) \cite{Buchholz:2009fk,Do:2009tg,Bartko:2010fk,Do:2013fk}.  And although it is unclear whether or not the B stars and the O/WR stars were formed in the same starburst, the radial profiles of these two groups exhibit a similar slope \cite{Do:2013fk}. Interestingly, the O/WR stars have a sharp inner edge at $R \sim$ 0.8", while the B stars continue inward toward Sgr A* \citeaffixed{Paumard:2006xd}{e.g.,}.

\begin{figure} \begin{center} \includegraphics[scale=0.35]{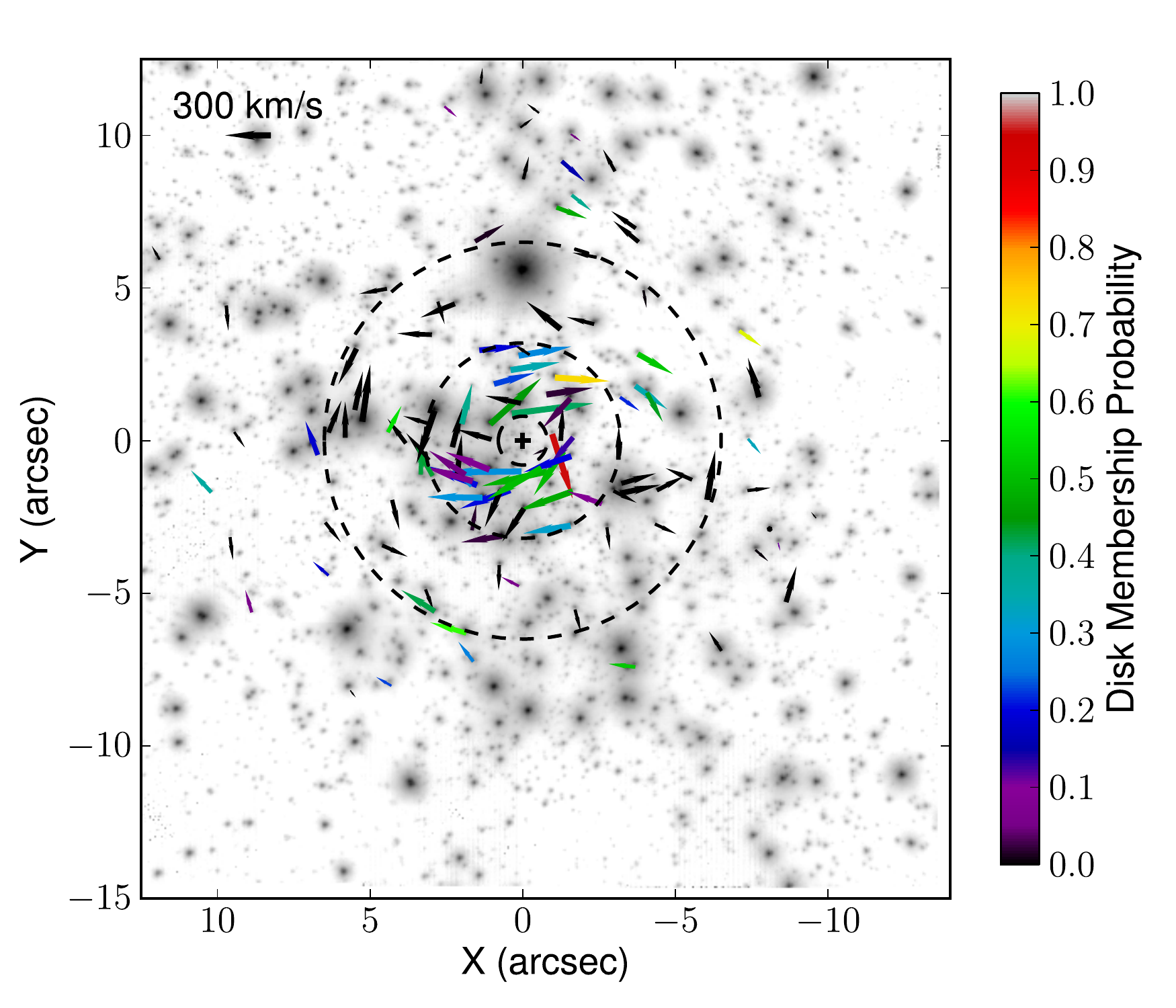} \caption{  Proper motion vectors of 116 young stars from \citeasnoun{Yelda:2014fk}.  The arrows are color-coded according to their disk membership probability.  Sgr A* is marked with a cross at the center and north is up and east is to the left. The dashed circles mark various radial extents from Sgr A* at $R$ = 0.8", 3.2", and 6.5". Coherent motion in the clockwise direction is clearly seen among many of the stars, the majority of which are within $R \sim$ 3.2".  (reproduced with permission from {\it The Astrophysical Journal}).  \label{fig:yngVels}}
\end{center}
\end{figure}

Coherent proper motions of the O and WR stars in the clockwise direction were first noted by \citeasnoun{Genzel:2000hc}. Many of these clockwise-rotating stars were later shown by \citeasnoun{Levin:2003kx} to be orbiting in a thin disk with half-opening angle $<$10$^{\circ}$. With the advent of adaptive optics, spectroscopic identification of more young stars down to $K\sim$16 (early B-type main sequence stars) was made possible, along with measurements of the stars' line-of-sight velocities. The increasingly larger samples and more precise kinematic information has led to improved and more detailed knowledge of many of the disk's properties.  The orientation of the orbital plane is ($i$, $\Omega$) = (125$^{\circ}$, 100$^{\circ}$)\footnote{We report the average inclination and angle to the ascending node from the given references.}, where $i$ is the inclination and $\Omega$ is the angle to the ascending node measured eastward of north \cite{Genzel:2003it,Paumard:2006xd,Lu:2009bl,Bartko:2009fq,Yelda:2012tw,Yelda:2014fk}. \citeasnoun{Bartko:2009fq} found that the disk's orientation changes as a function of radius from Sgr A*, although \citeasnoun{Yelda:2014fk} found no significant kinematic structures between 3.2"-6.5". The latter result suggests that the clockwise disk exists only in the innermost region of the Galaxy with a radial extent of approximately 0.15 pc and is surrounded by a more isotropically-distributed population of young stars.  Using Monte Carlo simulations of mock data, \citeasnoun{Yelda:2014fk} showed that the disk contains only $\sim$20\% of the 116 young stars in their sample (Figure \ref{fig:yngVels}), but includes some B-type main sequence stars.  The disk has an eccentricity of $e \sim$ 0.3 and a relatively steep surface density profile that scales as $\Sigma(R) \sim$ R$^{-2}$ \cite{Beloborodov:2006ff,Paumard:2006xd,Lu:2009bl,Bartko:2009fq,Bartko:2010fk}.

The discrepancy in the literature regarding the location of the outer edge of the disk and the dynamical properties of the non-disk members is in part due to the lack of acceleration information for the more distant stars and to the different assumptions used when defining disk membership \cite{Lu:2009bl,Bartko:2009fq,Bartko:2010fk,Yelda:2014fk}.  Without acceleration detections in the plane of the sky, a star's line of sight distance is unknown, making its orbit highly uncertain and dependent on assumptions. Early work using 2D positions and 3D velocities led to claims of a counterclockwise (CCW) disk that was nearly orthogonal to the clockwise system \cite{Genzel:2003it,Paumard:2006xd}. As kinematic measurements improved as a result of increased time baselines, larger fields of view, and higher precision astrometry and spectroscopy, the existence of a well-defined counterclockwise system has become less certain.  \citeasnoun{Bartko:2009fq} and \citeasnoun{Bartko:2010fk}, who conducted a Monte Carlo orbital analysis on $\sim$90 O and WR stars, reinterpreted the CCW system as a dissolving disk or streamer.  \citeasnoun{Lu:2009bl} and \citeasnoun{Yelda:2014fk}, on the other hand, added acceleration measurements and upper limits and found that the dynamical properties of stars not on the CW system were consistent with those of an isotropic population.  Increasing both the precision and time baseline of astrometric measurements will be critical for fully understanding the dynamical state of the stars at large radii.

Several theories have been put forward to explain the origin of the young, massive stars in the GC. The coherent motion of many of the stars may be indicative of {\it in situ} formation in a massive, gas disk that fragmented under its own self-gravity \cite{Levin:2003kx}. The stellar surface density is predicted to scale as $R^{-2}$ \cite{Lin:1987fv,Levin:2007bs}, consistent with observations within the CW disk plane \cite{Paumard:2006xd,Lu:2009bl,Bartko:2009fq}.  A slowly built-up gas disk would result in circular stellar orbits, which can then be excited to the present-day values of $e \sim$ 0.3 through two-body interactions if the mass function were top heavy \citeaffixed{Alexander:2007lh,Yelda:2014fk}{see \S\ref{sec:imf};}.  Alternatively, a stellar disk may result from the inward migration of a massive cluster whose stars are tidally stripped as it spirals inward under dynamical friction \cite{Gerhard:2001uq}.  However, to transport the stars to their present-day Galactocentric radii ($R \sim$ 0.04 pc) requires unrealistic cluster properties, including extremely high masses and densities, possibly even a central intermediate mass black hole \citeaffixed{Hansen:2003fk,Kim:2003fk,Kim:2004fu,Gurkan:2005fk,Berukoff:2006dz}{IMBH;}.  Furthermore, the stars stripped from an infalling cluster will follow a shallow surface density profile scaling as $R^{-0.75}$ \cite{Berukoff:2006dz}, in contrast with the observations. Thus, the cluster inspiral scenario is difficult to reconcile with the observations to date, and it is therefore more likely that the young stars formed {\it in situ}.

If the massive stars in the central parsec formed together as many have suggested \cite{Paumard:2006xd,Lu:2009bl,Bartko:2009fq,Yelda:2014fk}, some dynamical evolution must have occured that led to the complex present-day configurations.  Assuming a single-disk origin, for example, requires some dynamical mechanism(s) that can excite the orbits such that up to 80\% \cite{Yelda:2014fk} of the stars can no longer be associated with the CW disk. Massive perturbers have been invoked, including a theorized inward-migrating IMBH \cite{Yu:2007fu} and the observed circumnuclear disk located at $R \sim$ 1.5 pc \citeaffixed{Christopher:2005fk,Subr:2009kl}{CND;}.  In the case of the CND, differential precession can lead to a configuration that is similar to what is observed (i.e., a compact stellar disk at the innermost radii and stars with large inclinations to the disk at large radii). In light of the possibility of the existence of a second, counterclockwise disk, \citeasnoun{Lockmann:2009ye} also considered the effects of two highly inclined disks of different masses. They found that the mutual interaction of the two structures would lead to the ultimate destruction of the lower mass (CCW) system within 5 Myr, leaving no observational signatures of that disk today.

\subsection{Enclosed Mass Estimate from the Kinematics of the Young Stars}
\label{sec:LM}
Knowledge of the young stars' proper motions allows for an estimate of the enclosed mass using the Leonard-Merritt (LM) mass estimator \cite{Leonard:1989fk}. Following \citeasnoun{Schodel:2009zr} and using data from \citeasnoun{Yelda:2014fk}, we compute the enclosed mass as a function of distance from Sgr A* (Figure \ref{fig:LM_lu13KLF}, left panel). The distance to the GC is fixed at 8 kpc.  We find that the mass begins to plateau at $\sim$0.3\,pc. The proper motions of the young stars yield a mass of $M_{enc}$ = 3.8 $\pm$ 0.2 $\times$10$^6$ M$_{\odot}$. The uncertainty is estimated by calculating the LM mass estimator on N subsamples, each of which have a similar radial distribution.

It is encouraging that the proper motions of the young stars alone give an enclosed mass that is fully consistent with those obtained from the orbital analyses of short-period stars for a GC distance of 8\,kpc \citeaffixed{Ghez:2008fk,Gillessen:2009qe}{e.g.,}. However, as Fig.\,\ref{fig:mass} shows, the {\it stellar mass} enclosed within $\sim$$0.3-0.5$\,pc may already be a few $0.1$ up to $1\times10^{6}$\,M$_{\odot}$. Therefore, we should expect no leveling off  of the enclosed mass beyond $\sim$0.3\,pc. Two possible explanations for the levelling off are: (a) The stellar mass in the central 0.1\,pc may be smaller than what is  shown in Fig.\,\ref{fig:mass}. In this context we recall that the model shown \citeaffixed{Feldmeier:2014kx}{for more detials, see discussion and Fig.\,17 in} is tied to the surface brightness profile of the cluster, which may not be a good mass tracer at small distances from Sgr\,A*. (b) The tracer population of massive young stars may not be complete at distances $>0.3\,$pc from Sgr\,A*. Indeed, the spectroscopic coverage of the field around Sgr\,A* becomes more incomplete and variable in sensitivity at greater distances from Sgr\,A* \cite{Bartko:2010fk,Do:2013fk}.

\subsection{Stellar Mass Function}
\label{sec:imf}
In addition to kinematics, the stellar mass function is important for constraining theories of star formation in the GC.  Given the hostile conditions in the region, one might expect the initial mass function (IMF) to differ from the standard stellar mass distribution seen in normal Galactic star forming regions \citeaffixed{Salpeter:1955qo}{$dN/dm \propto m^{-\alpha}$, $\alpha = 2.35$;}.  Indeed, observations have consistently shown this to be the case.  {\it Chandra} observations of the GC revealed that the X-ray emission from low mass stars, whose surfaces have high magnetic activity, is an order of magnitude lower than expected for a canonical IMF \cite{Nayakshin:2005ve}. Using adaptive optics spectroscopic observations, \citeasnoun{Paumard:2006xd} constructed a K-band luminosity function consisting of the brightest, most massive stars ($K <$ 13, $M >$ 20 M$_{\odot}$) and claimed a top-heavy mass function. Later, \citeasnoun{Bartko:2010fk} extended the \citeasnoun{Paumard:2006xd} observations by including deeper spectroscopy down to main sequence B stars ($K <$ 16, $M >$ 10 M$_{\odot}$) and found an extremely top-heavy mass function with slope $\alpha$ = 0.45 $\pm$ 0.3.

\begin{figure} \begin{center} \includegraphics[scale=0.45]{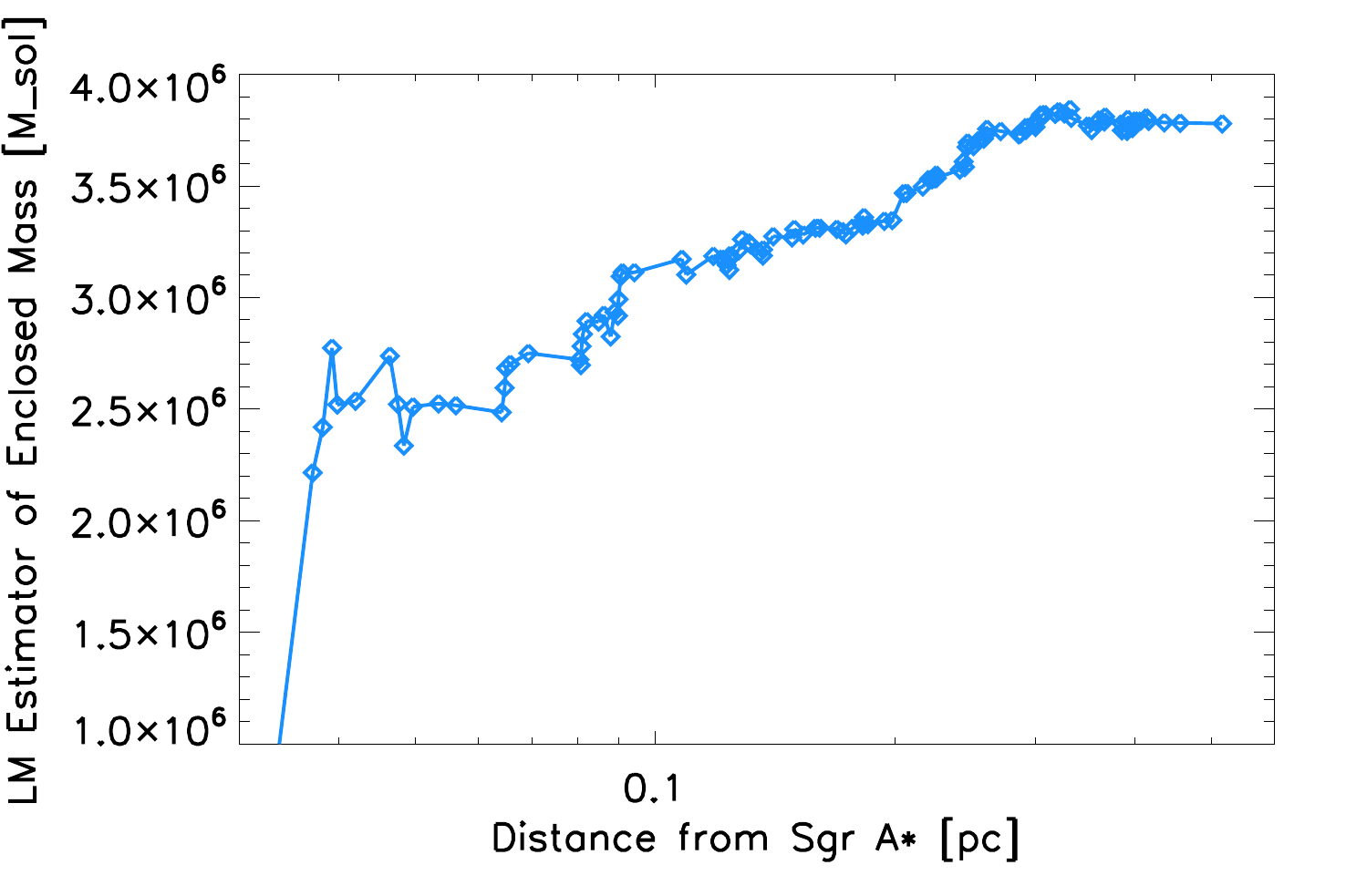} \includegraphics[scale=0.2]{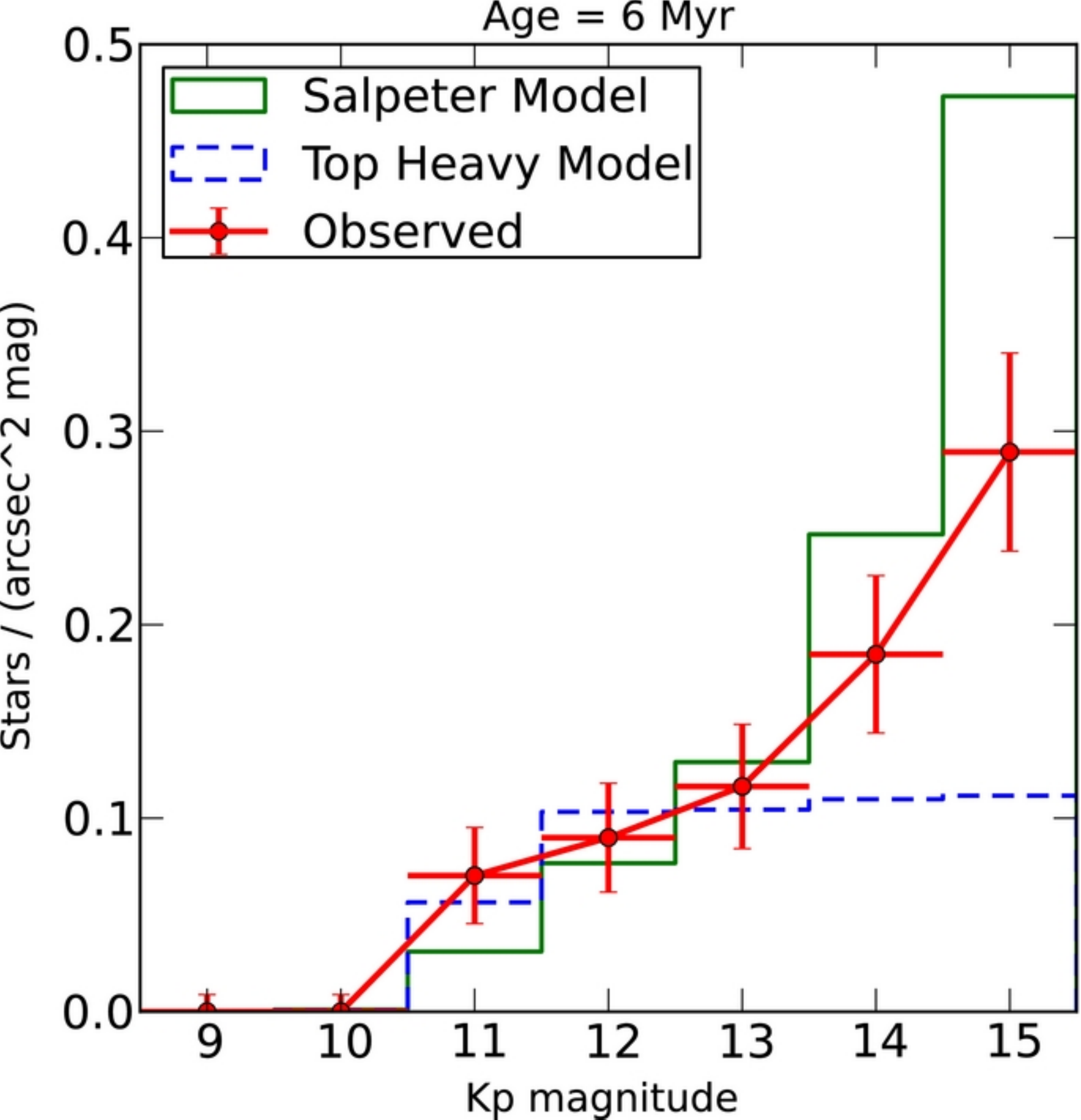} \caption{ ({\it Left:}) Leonard-Merritt (LM) mass estimator as a function of maximum projected distance based on proper motion measurements of the young stars in \citeasnoun{Yelda:2014fk}.  Each data point represents the LM estimate of the enclosed mass using stars interior to that point. The estimated mass plateaus at $\sim$3.8$\times$10$^6$ M$_{\odot}$ at $R >$ 0.3 pc.  ({\it Right:}) Observed $Kp$-band luminosity function of the young stars in the central parsec, as shown in Figure 1 of \citeasnoun{Lu:2013fk}. The observed magnitudes ({\it red}) were measured in \citeasnoun{Do:2013fk} and were corrected for differential extinction. The model KLFs for a Salpeter (with slope $\alpha$ = 2.35; {\it green}) and a top-heavy (with slope $\alpha$ = 0.45; {\it blue}) mass function are overplotted.  The observations suggest a mass function that is slightly more top-heavy than Salpeter  (reproduced with permission from {\it The Astrophysical Journal}).  }
\label{fig:LM_lu13KLF} 
\end{center}
\end{figure}

Recently, \citeasnoun{Do:2013fk} and \citeasnoun{Lu:2013fk} conducted a more robust statistical analysis using a Bayesian inference methodology, which included prior information on the underlying population and extensive simulations of synthetic stellar clusters. Their observed $Kp$-band luminosity function (KLF) along with model KLFs for a Salpeter and a top-heavy mass function are shown in the right panel of Figure \ref{fig:LM_lu13KLF}. Their results are consistent with a moderately top-heavy IMF, with slope $\alpha$ = 1.7 $\pm$ 0.2, which is inconsistent with the \citeasnoun{Bartko:2010fk} work. Aside from the differences in the statistical methodologies, these studies also differed in the sample used. The \citeasnoun{Bartko:2010fk} sample consisted of many stars perpendicular to the clockwise disk and did not include the innermost S-stars.  In contrast, the \citeasnoun{Do:2013fk} and \citeasnoun{Lu:2013fk} samples extended along the clockwise stellar disk and included all observed young stars, including the S-star cluster. However, when excluding the central arcsecond stars from the latter work, neither the shape of the K-band luminosity function nor the slope of the IMF are significantly changed.  Furthermore, the number of low mass stars (0.5-3 M$_{\odot}$) predicted based on the \citeasnoun{Lu:2013fk} IMF slope is a factor of 4-15 times lower than expected for a Salpeter IMF, a result that is in agreement with the X-ray observations of \citeasnoun{Nayakshin:2005ve}. The GC therefore represents one of a few known regions to have a non-canonical initial mass function.

As a  caveat we add, however, that this non-canonical, top-heavy IMF may only apply to the most recent star formation event at the GC.  The   IMF of the older stars in the MWNSC is subject to ongoing research.
While some work provides support for a long-standing top-heavy IMF at the GC \citeaffixed{Maness:2007sj}{e.g.,}, other studies on the older stellar population in the central parsec of the NSC have come to the conclusion that its properties are consistent with a standard IMF \citeaffixed{Pfuhl:2011uq,Lockmann:2010fk}{e.g.,}.

\section{Stellar dynamics near the central black hole and upcoming tests of GR} 
\label{sec:GR}

\begin{figure}
\centering
\includegraphics[width=9cm]{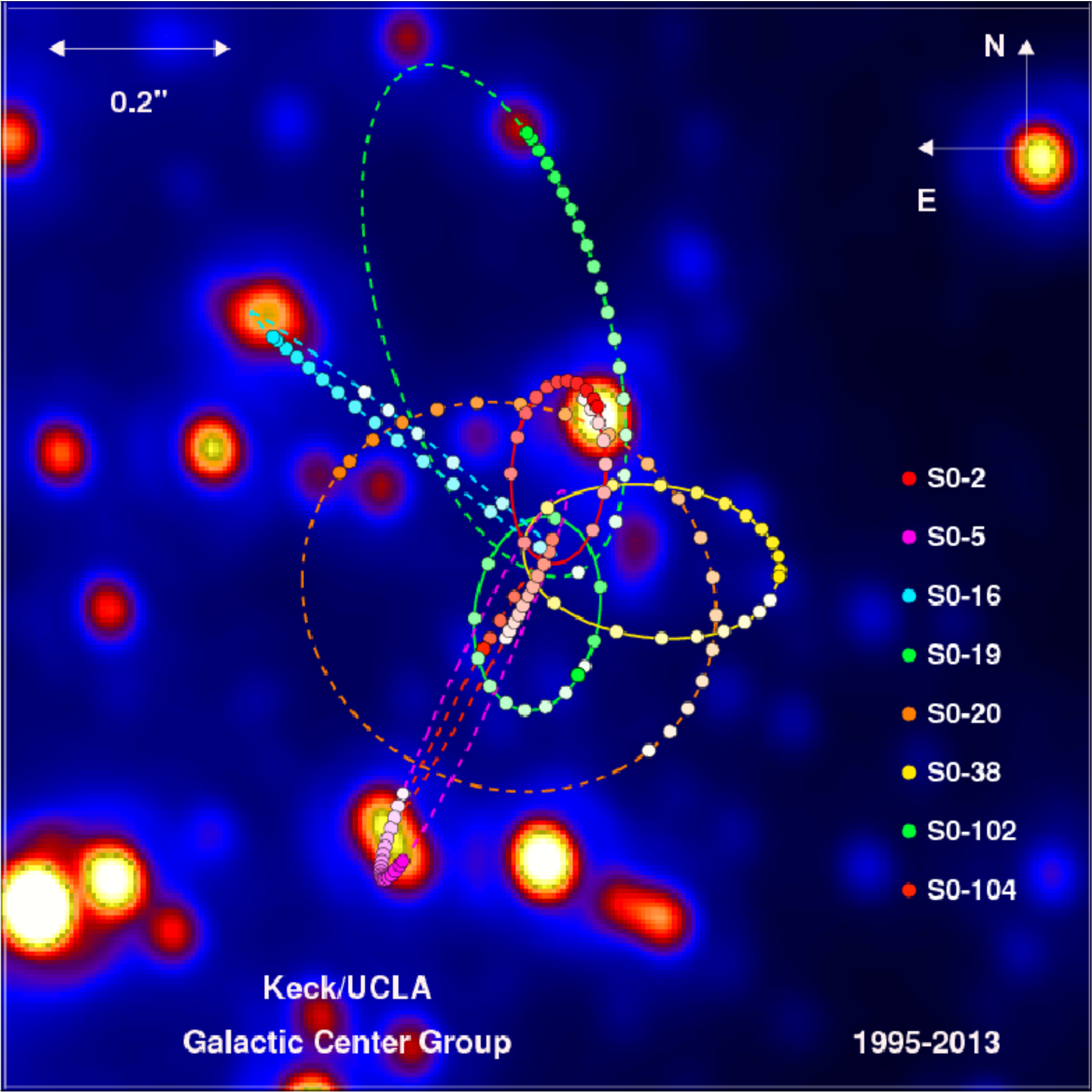}
\caption{\label{Fig:centImg}  Orbits of the best known short-period S-stars at the GC. The three stars plotted with solid lines (S0-2, S0-102, S0-38) have orbital periods less than 20 years and have been traced for a whole orbit \cite[Boehle et al., in prep.]{Ghez:2008fk,Meyer:2012fk}.  }
\end{figure}

In this section, we will turn our attention to the population of stars closest to the central black hole: the so-called S-star cluster (Fig.\,\ref{Fig:centImg}). These are the stars within a central radius of 0.8'' (0.03 pc). They are primarily B-type main sequence stars (at least the brighter ones with K $<$ 16 for which spectroscopy is currently feasible)  and constitute a dynamically distinct population, since compared to the O/WR/B stars further out, they are isotropically distributed and highly eccentric with a mean eccentricity of 0.8 \cite{Schodel:2003qp,Ghez:2005fk,Gillessen:2009qe}. The eccentricity distribution has been reported to be consistent with a thermal distribution, $n(e) \sim e$, albeit favoring somewhat higher eccentricities \cite{Gillessen:2009qe}. While the dynamical properties of the S-stars are clearly different from the B-star population beyond a central radius of 0.8'', it is unclear whether they are an inner extension of the starburst that is manifested in the stellar disk of young stars (see previous section) or have been formed in a distinct star formation event. Alternatively, they may have been formed elsewhere and at other times in the MWNSC and been deposited at their current locations through tidal capture of binaries, accompanied by the ejection of hypervelocity stars \citeaffixed{Hills:1988zh,Gould:2003dq,Perets:2007ud}{e.g.,}. As noted by \citeasnoun{Do:2013fk}, the relevant part (K $>$ 14.0) of the S-stars luminosity function is consistent with the luminosity function for stars at $>$ 1'' of the same magnitude range, which might or might not point to a common origin.

The apparent young age, the proximity to the black hole, and the distinct kinematic properties pose the question of how the S-star cluster was created. It seems clear that the stars have not formed {\it in situ} but have been brought in either as a member of a binary system on a radial orbit or through migration from the stellar disk. In the binary capture mechanism, a binary star gets disrupted in an interaction with the black hole that leaves one star in a highly eccentric, tight orbit while the other star escapes the nuclear star cluster \cite{Hills:1988zh}. The high eccentricities of the orbiting stars then need to almost thermalize to match the observations. Most recently, \citeasnoun{Antonini:2013il} found that this is feasible quickly enough, but only when the nuclear star cluster has a cusp, which is not observed in the population of old giants (see Section\,\ref{sec:cusp}). A constant supply of low angular momentum orbits for binaries is hypothesized to be caused by massive perturbers at distances $>$ 1pc \cite{Perets:2007ud}. While the question is not settled, the binary disruption scenario is currently preferred over a disk-migration model \cite{Levin:2007bs}, in part because it offers an explanation of the hyper-velocity stars that escape the Galaxy and some of which may have originated at the Galactic Center \citeaffixed{Brown:2009ly}{e.g.}.

The most prominent member of the S-stars is S0-2/S2, a K=14 mag star in a 16 year period around the central black hole. It was co-discovered by the UCLA  and MPE groups in the mid-1990's and has been key to establishing both the presence and characteristics of our Galaxy's central black hole \cite{Ghez:1998ad,Ghez:2003fk,Ghez:2005fk,Ghez:2008fk,Eckart:1997jl,Schodel:2002zt,Gillessen:2009nx,Gillessen:2009qe}. Its relatively bright magnitude and short orbital period resulted in it being the first star which has been monitored for a complete orbit, thereby dominating the stellar probes that measure the central gravitational potential.  The latest published values for the mass of and distance to the Galactic black hole as inferred from S0-2's orbit are $M = 4.3 \pm 0.4 \times 10^6 M_\odot$, $R_0 = 7.9 \pm 0.4$ kpc as reported from Keck Observatory measurements \cite{Meyer:2012fk}\footnote{The cited paper quotes the values for the overall best fit, whereas here we give the expectation and standard deviation for the marginalized, one-dimensional distributions.}, and $M = 4.3 \pm 0.4 \times 10^6 M_\odot$, $R_0 = 8.3 \pm 0.4$ kpc as reported from VLT observations \cite{Gillessen:2009qe}.
S0-2 itself appears to be a main-sequence B0-2.5 V star with a low rotation velocity and a slight He enrichment \cite{Ghez:2003fk,Martins:2008fe}.

Short orbital periods are key for the accurate and precise determination of the Keplerian elements and the central potential. An insufficiently covered orbital phase leads to a bias in the posterior of the orbital elements. \citeasnoun{Lucy:2013gb} finds that at least 40\% of the orbit needs to be covered with observations to ensure an unbiased estimate. Until recently S0-2 was the only known star with an orbital period of less than 20 years. Then, the multi-year AO observations enabled the discovery of S0-102, a K=17 mag star with a period of a mere 11.5\,years \cite{Meyer:2012fk}. Additionally, S0-38/S38 could recently be traced back for almost a complete orbit (Boehle et al., in prep.), which solidified the initially reported orbital period of 19 years \cite{Gillessen:2009qe}.  These additional short-period stars will cross-check the results from S0-2 and further constrain the central potential when used in a combined fit (Boehle et al., in prep.).

With the astrometric AO-assisted imaging programs continuing, the next frontier in the determination of stellar orbits is the detection of post-Keplerian effects. Carrying  out these measurements offers the unique opportunity to test General Relativity (GR) -- the least tested of the four fundamental forces of nature -- in an unexplored regime. S0-2 probes gravity regimes that are of magnitude $\epsilon\approx GM/(Rc^2) \approx 10^{-4}$. Here, $G$ is the gravitational constant, $M$ the mass, $R$ the distance, and $c$ the speed of light. This is two orders of magnitude stronger than solar system scales, where Einstein's theory of general relativity has so far passed all tests with flying colors. The gravitational fields that have been probed in tests using double neutron stars such as the Hulse-Taylor binary pulsar are of the same magnitude, because the masses and separation of the neutron stars are comparable to the mass and radius of the Sun (note, however, that this applies to the inter-body potential only and not the strong-field internal gravity of the neutron stars).

Specifically, future measurements of the short-period stars afford the opportunity to probe the structure of space-time in a gravitational (inter-body) potential that is 100 times stronger and on a mass scale that is 400,000 times larger than any other established existing test, providing probes of GR that many theorists have long anticipated \citeaffixed{Jaroszynski:1998mb,Fragile:2000cq,Rubilar:2001fk,Loeb:2003kh,Weinberg:2005fk,Zucker:2006uq,Kraniotis:2007rq,Will:2008fc,Merritt:2010vn,Angelil:2010qc,Angelil:2010kx,Angelil:2011uq,Iorio:2011bd,Sabha:2012ud,Psaltis:2012if,Psaltis:2013kc}{e.g.}. The first opportunity to detect one of the effects of relativity arises during the next closest approach of S0-2. During this passage, predicted to occur in the summer of 2018, the difference in the radial velocity between a Newtonian description of the star and photon orbit and a relativistic one will peak at 200\,km/s \cite{Zucker:2006uq}. This value refers to the difference between a pure Newtonian calculation of S0-2's redshift and a relativistic one. This difference is a function of orbital phase, since the star's velocity and distance to the BH is a function of orbital phase. It is in roughly equal parts due to the special relativistic Doppler shift and the gravitational redshift. To measure these relativistic deviations, astrometric and spectroscopic measurements are required over S0-2's entire orbit to determine the Keplerian orbital elements (13 in total) with sufficient precision to ensure that the 2018 measurements are sensitive to the relativistic terms. 

\begin{figure}
\centering
\includegraphics[width=8cm]{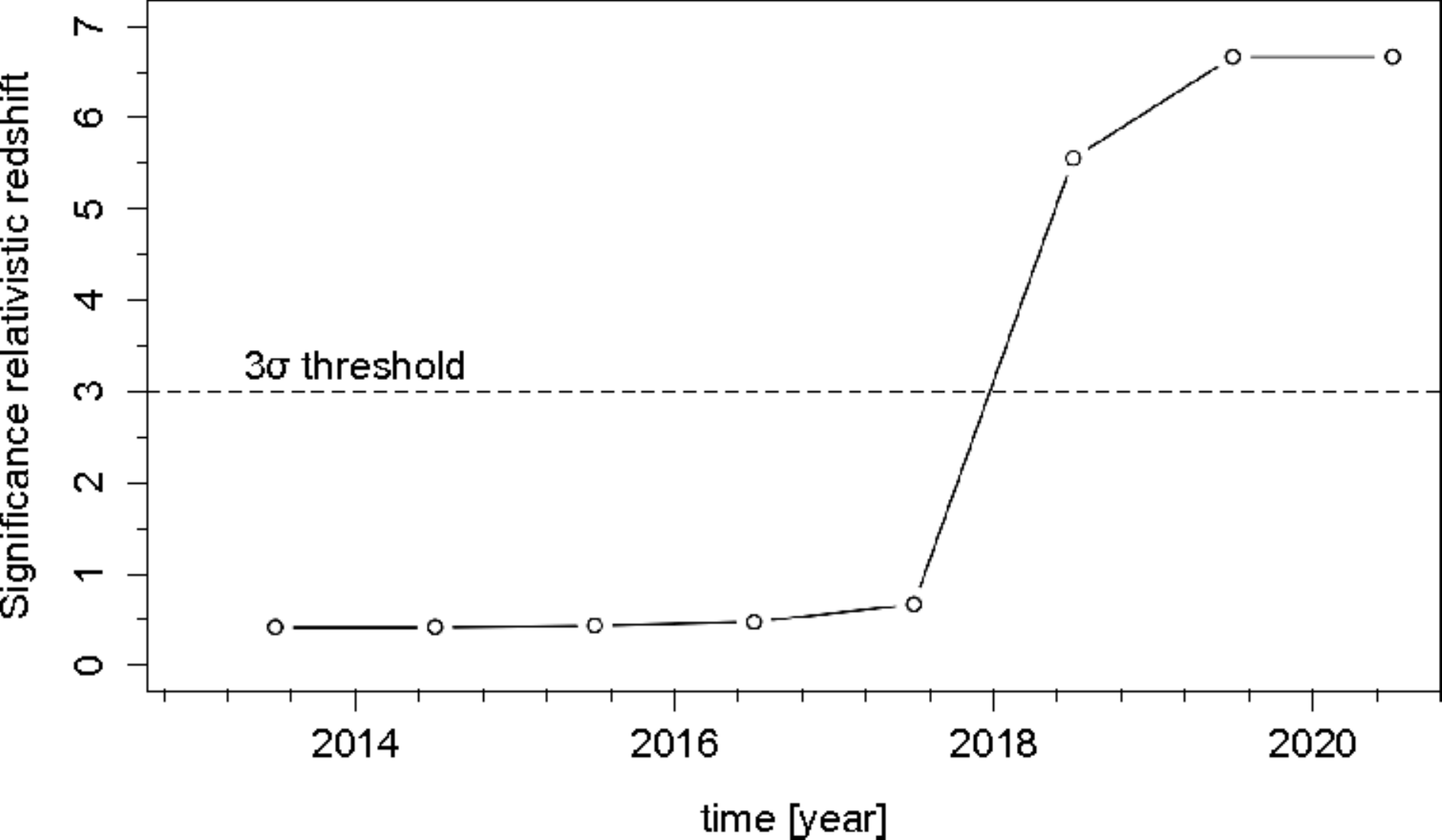}
\caption{\label{Fig:Rel}  The sensitivity to the relativistic redshift of S0-2 as a function of time, based on the current observational approach and typical uncertainties. After S0-2 passes its closest approach in 2018, a 6.5$\sigma$ result is expected. }
\end{figure}

The current spectroscopic uncertainty of the line-of-sight velocity of S0-2 is 20 km/s. In the astrometric domain, there are three error terms that need to be considered: the centroiding error which determines the precision with which a star can be located on the detector (in pixel coordinates), a purely empirical additive error that is added in quadrature to ensure $\chi^{2}$ statistics, and an alignment error that is a consequence of the transformation of each epoch into a common reference system \cite{Clarkson:2008fk,Yelda:2010sl,Yelda:2014fk}. We have used typical past values for the centroiding error ($0.12$\,mas) and the additive error ($0.1$\,mas). The alignment error is different for each epoch (zero for the reference epoch) and depends on the the number of wide field maser mosaics taken that are used to tie the infrared observations into a radio reference frame. We have estimated the alignment error slope to be 20\,$\mu$as/yr in 2018. The error slope determines the alignment error for a given epoch using the number of years that epoch is away from the reference epoch. Assuming negligible systematic errors and an observing cadence of 2 astrometric measurements per year (one early in the season and one late) and 2 radial velocity measurements per year (except in 2018 where 10 measurements have been assumed), we conclude that a 6.5$\sigma$ test of the relativistic redshift and therefore Einstein's Equivalence Principle by the end of 2019 based on S0-2's measurements alone can be anticipated (see Fig.\,\ref{Fig:Rel}).

With the steady increase of the time baseline and the advent of next-generation facilities like the giant segmented mirror telescopes (with a diameter of $\sim$30 m) or sensitive instrumentation on  near-infrared interferometers, such as VLTI/GRAVITY \citeaffixed{Gillessen:2010kx}{e.g.}, it is expected that more short period stars around the black hole will be detected and accurately tracked on their orbits. Taken together with the already known short-period stars, deviations from Keplerian orbits will be visible in the spectroscopic as well as imaging domain for several stars. For astrometry, the most important, lowest-order post-Keplerian effect is the precession of the periapse, which has two causes leading to opposite effects: a prograde precession as described by the GR equations of motion in the Schwarzschild metric, and a retrograde precession as described by the Newtonian equations of motion in an extended, spherically symmetric mass configuration. To disentangle both effects, the periapse precession needs to be detected in several stars. Such a detection would constitute a very important measurement, as this (1) tests the specific metric form of General Relativity in an unprecedented regime and can therefore distinguish between different metric theories of gravity, and (2) probes the distribution of dark stellar remnants around the black hole and thereby tests fundamental models of galaxy evolution and N-body dynamics \cite{Rubilar:2001fk,Weinberg:2005fk,Merritt:2010vn}. Looking to higher order effects and therefore further into the future, \citeasnoun{Will:2008fc} noted that stellar orbits offer the possibility to measure the black hole's quadrupole moment and thereby to test the no-hair theorem, which states that the quadrupole moment is uniquely described by the angular momentum and mass of the black hole.

How feasible are the astrometric measurements of GR effects in stellar orbits around the central black hole? From an observational perspective, the stable construction of an absolute astrometric reference frame is one of the biggest challenges. We note that our calculation of the significance of the measurement of the relativistic redshift presented above assumes that the description of a linear drift captures the apparent reference frame behavior sufficiently. The key point is that in this case the estimation of the Keplerian orbital parameters is unbiased and therefore such is the relativistic redshift measurement. For the observation of the precession of the periapse, however, the signal is in the astrometric domain (in contrast to the radial velocity spectroscopic measurements), and a drift cannot be easily disentangled from a periapse shift. This is why additional constraints on the stability of the reference frame are required. As inferred from S0-2's orbit, the current stability of the reference frame is $\sim0.5$\,mas/yr, whereas the detection of the GR periapse precession of S0-2 requires a stability of $<0.05$ mas/yr. Recent simulations suggest that the spatial variability of the point spread function and the camera distortion are the main systematic causes of this reference frame drift (Meyer et al., in prep.). At UCLA, projects are already underway to correctly model these effects for current systems \citeaffixed{Fitzgerald:2012hs}{PI A. Ghez, see}, and next-generation AO systems like  multi-conjugate AO will provide a hardware solution \citeaffixed{Fitzgerald:2012hs}{e.g.}.  The future therefore seems to be bright.

\section{Summary and outlook}

It appears that the Milky Way's nuclear star cluster is indeed a typical specimen of its kind. This is reflected by its overall properties, with its size, mass, luminosity, rotation, and stellar population being very similar to what we observe in external systems. Although the data appear to suggest that the MWNSC is relatively massive for its size, it is not clear whether this is indicative of some astrophysical anomaly or simply a consequence of the intrinsic scatter of NSC masses and, potentially, of observational bias.

Having established that the MWNSC is representative for its class, we are in a privileged position to study phenomena that are thought to occur generally in NSCs at a level of detail that will always be unachievable in extragalactic systems. When studying the MWNSC, it is, however, always important to know about the observational limitations and understand how specific data have been obtained and corrected for any biases before drawing any conclusions. There are, on the one hand,  observational obstacles that  will be overcome by future instruments and telescopes. This concerns primarily the problem of crowding that can be addressed with improved AO systems and larger telescopes. But, on the other hand, there are also observational constraints that are fundamental in nature, mainly the extreme and variable interstellar extinction and the impossibility to perform any reasonably sensitive observations of the stellar population shortward of about $1.3\,\mu$m. We will therefore probably not be able to observe significantly sub-solar-mass stars or objects such as white dwarfs near the GC, even with a telescope of the 30-40m-class.

Current observational work has shown that the MWNSC apparently displays no stellar cusp around the central MBH, when we focus on the bright late-type stars and the brightest RC stars. This may reflect the actual state-of-affairs, i.e.\ the bright late-type stars may be representative for the structure of the entire cluster \citeaffixed{Merritt:2010ve}{e.g.,}, but it may also be only an {\it apparen}t lack of a cusp, in case the envelopes of giants have been destroyed \citeaffixed{Dale:2009ca,Amaro-Seoane:2014fk}{e.g.,}. The jury is still out on this question. Some progress can be expected in the coming years from more and better spectroscopic or, possibly, photometric data. The case will definitely be closed once diffraction limited observations with extremely large telescopes become possible. With the current timeline for projects such as the TMT or the E-ELT, this can be expected to happen in the first half of the 2020s.

The young, massive stars near Sgr\,A* can teach us much about how star formation can proceed in a NSC. The case of the MWNSC provides strong evidence that {\it in situ} star formation can take place in NSCs, even close to a massive black hole, probably  in a formerly existing dense gas disk. There is reasonable  evidence that the IMF was top-heavy in the latest star formation event, a possible consequence of the extreme environment.  A top-heavy IMF may lead to the production of a large number of stellar-mass black holes over the life time of a cluster and thus boost event rates for EMRIs.  But, as a word of caution, the top-heavy IMF appears to apply only to the most recent starburst. The bulk of the stars of the MWNSC may have formed with a more normal IMF \cite{Pfuhl:2011uq,Lockmann:2010fk}. 

One of the most exciting aspects of the MWNSC is that we can observe the orbits of stars around the central BH. These measurements do not only allow us to obtain ever more accurate estimates of the mass and distance of Sgr\,A*, but also bear the promise of upcoming tests of GR in a so far unexplored regime. So far, three stars with orbital periods $<20$\,yr have been detected. Improved technology (e.g., adaptive optics) and methodology (e.g., PSF modeling), as well as the increasing time baseline of astrometric and spectroscopic measurements will probably result in the detection of  more short-period stars in the coming years, thus leading to an ever more precise determination of the mass, distance, position, and motion of Sgr\,A*. The next closest approach of the star S2/S0-2 to Sgr\,A*, predicted to occur in summer 2018, will be a milestone in GC research. It can be expected that this event will allow the involved research groups to perform the first highly significant tests of General Relativity in the neighbourhood of Sgr\,A*, in particular tests of Einstein's Equivalence Principle. Tests of the {\it metric} of General Relativity will be harder to perform. They will require the use of 30-40m class telescopes over one or several decades, depending on the number of stars on short-period orbits around Sgr\,A* that will be detected with such future facilities. Here, we can roughly consider a time horizon of 15-20 years. The planned commissioning of the GRAVITY instrument at ESO's Very Large Telescope Interferometer may offer the prospect that such a feat can be performed earlier. If GRAVITY meets its design goals and starts operating as foreseen in 2015 -- and depending on (1) the unknown number of faint stars and their orbits at distances of less than about two milli parsecs from Sgr\,A*, and (2) whether the astrometric reference frame of the extremely small field-of-view of GRAVITY can be kept sufficiently stable -- it may be possible to measure the metric of GR near Sgr\,A* already in the early 2020s.

\ack

RS acknowledges support by grant AYA2010-17631  of the Spanish Ministry of Economy and Competitiveness. The research leading to these results has received funding from the European Research Council under the European Union's Seventh Framework Programme (FP/2007-2013) / ERC Grant Agreement n.\ 614922. NN and AF acknowledge support by the Excellence Cluster Universe. LM and SY acknowledge support from NSF grant AST-0909218.

\section*{References}
\bibliography{/Users/rainer/Documents/BibDesk/BibGC}

\end{document}